\documentclass[runningheads]{llncs}
\usepackage{ialgo2}
\usepackage{version} 
\usepackage[hidelinks]{hyperref}
\hypersetup{colorlinks=true,linkcolor=blue,citecolor=blue,urlcolor=blue}
\usepackage{orcidlink}
\title{Formalizing the Notions of \\
       Non-Interactive and Interactive Algorithms}
\author{C.A. Middelburg\,\orcidlink{0000-0002-8725-0197}}
\institute{Informatics Institute, Faculty of Science, University of
           Amsterdam \\
           Science Park~900, 1098~XH Amsterdam, the Netherlands \\
           \href{mailto:C.A.Middelburg@uva.nl}{C.A.Middelburg@uva.nl}}
  
\titlerunning{Formalizing the Notions of
              Non-Interactive and Interactive Algorithms}
\authorrunning{C.A. Middelburg}

\begin{document}
\maketitle

\begin{abstract}
An earlier paper gives an account of a quest for a satisfactory 
formalization of the classical informal notion of an algorithm.
That notion only covers algorithms that are deterministic and 
non-interactive.
In this paper, an attempt is made to generalize the results of that 
quest first to a notion of an algorithm that covers both deterministic 
and non-deterministic algorithms that are non-interactive and then 
further to a notion of an algorithm that covers both deterministic and 
non-deterministic algorithms that are interactive.
The notions of an non-interactive proto-algorithm and an interactive 
proto-algorithm are introduced. 
Non-interactive algorithms and interactive algorithms are expected to be 
equivalence classes of non-interactive proto-algorithms and interactive 
proto-algorithms, respectively, under an appropriate equivalence 
relation.
On both non-interactive proto-algorithms and interactive 
proto-algorithms, three equivalence relations are defined.
Two of them are deemed to be bounds for an appropriate equivalence 
relation and the third is likely an appropriate one.
\begin{classcode}
F.1.1, F.1.2, F.2.0 
\end{classcode}
\end{abstract}

\section{Introduction}
\label{sect-intro}

In~\cite{Mid24a} an account is given of a quest for a satisfactory 
formalization of the notion of an algorithm that is informally 
characterized in standard works from the mathematical and computer 
science literature such as~\cite{Kle67a,Knu97a,Mal70a,Rog67a}.
The notion of an algorithm in question is generally considered the 
original notion of an algorithm. 
Because several generalizations of this notion have emerged, it is now 
often referred to as the notion of a classical algorithm.
Classical algorithms are known for their deterministic and 
non-interactive nature.

Non-deterministic algorithms, introduced in~\cite{Flo67a}, are widely 
used as a starting point for developing deterministic backtracking 
algorithms.
Moreover, they play an important role in the field of computational 
complexity.
In this paper, an attempt is made to generalize the results of the 
above-mentioned quest to a notion of an algorithm that covers both 
deterministic and non-deterministic algorithms that are non-interactive.
The term non-interactive algorithm is coined for this notion of an
algorithm.
Interactive algorithms, which emerged due to the advent of interactive 
computation, are currently widespread.
In this paper, an attempt is made to generalize the results of the 
above-mentioned quest further to a notion of an algorithm that covers 
both deterministic and non-deterministic algorithms that are 
interactive.

Interactive algorithms are mainly described with sentences like 
``An interactive algorithm is an algorithm that can interact with the 
environment in which it is applied''.
However, there is no consensus on which properties characterize  
interactive algorithms well.
In~\cite{BG06a}, a specific view on the nature of interactive algorithms 
is discussed in detail, culminating in a characterization of interactive 
algorithms by a number of postulates.
The view concerned is the only one found in the computer science 
literature so far. 
Some of its details are based on choices whose impact on its generality
is not clear.

Despite the fact that interactive algorithms are currently widespread, 
no other work aimed at a satisfactory formalization of the notion of an 
interactive algorithm has been reported in the computer science 
literature.
This state of affairs motivated me to start a quest for a satisfactory 
formalization of the notion of an interactive algorithm.
The main goals of this quest are to provide a framework for studying 
complexity-theoretic issues concerning interactive algorithms and to
provide a semantic basis for languages to describe interactive 
algorithms.

Due to the advent of interactive computation, several related models of 
interactive computation have been proposed.
They are based on variants of Turing machines, to wit interactive Turing 
machines~\cite{LW01a}, persistent Turing machines~\cite{GSAS04a}, and 
reactive Turing machines~\cite{BLT13a}.
In this paper, the starting point for the formalization of the notion of 
an interactive algorithm is a characterization of the notion by 
properties suggested by those models of interactive computation.

In~\cite{Mid24a}, first the notion of a classical proto-algorithm is 
introduced and then three equivalence relations on classical 
proto-algorithms are defined.
The thought is that classical algorithms are equivalence classes of 
classical proto-algorithms under an appropriate equivalence 
relation.
Two of the three equivalence relations defined give bounds for an 
appropriate equivalence relation and the third one is likely an 
appropriate one.
In this paper, the same approach is followed for non-interactive 
algorithms and interactive algorithms.

This paper is organized as follows.
First, the basic notions and notations used in this paper are introduced 
(Section~\ref{sect-preliminaries}).
Next, intuitive characterizations of the notion of a non-interactive 
algorithm and the notion of an interactive algorithm are given by 
properties that are considered to belong to the most important ones of 
these notions (Section~\ref{sect-algo-informal}).
After that, the formal notion of a non-interactive proto-algorithm is 
introduced and three relevant equivalence relations on non-interactive 
proto-algorithms are defined (Section~\ref{sect-noninter-algo}).
Then, the formal notion of an interactive proto-algorithm is introduced 
and three relevant equivalence relations on interactive proto-algorithms 
are defined (Section~\ref{sect-inter-algo}).
Thereafter, the connection between non-interactive proto-algorithms and
interactive proto-algorithms is addressed (Section~\ref{sect-versus}).
Finally, some concluding remarks are made 
(Section~\ref{sect-conclusions}).

In this paper, an attempt is made to generalize the work on the 
classical notion of an algorithm presented in~\cite{Mid24a}. 
In Sections~\ref{sect-algo-informal} and~\ref{sect-noninter-algo}, this 
has led to text overlap with~\cite{Mid24a}.

\section{Preliminaries}
\label{sect-preliminaries}

In this section, the basic notions and notations used in this paper are 
introduced.

The notion of a non-interactive proto-algorithm and the notion of an
interactive proto-algorithm will be formally defined in 
Sections~\ref{subsect-noninter-algo-proto} 
and~\ref{subsect-inter-algo-proto}, respectively, in terms of three 
auxiliary notions. 
The definition of one of the auxiliary notions is based on the 
well-known notion of a rooted labeled directed graph.
However, the definitions of this notion given in the mathematical and 
computer science literature vary. 
Therefore, the definition that is used in this paper is given.

\begin{definition}
\sloppy
A \emph{rooted labeled directed graph} $G$ is a sextuple 
$(V,E,\LBLv,\LBLe,l,r)$, where:
\begin{itemize}
\item
$V$ is a non-empty finite set, whose members are called the 
\emph{vertices} of $G$;
\item
$E$ is a subset of $V \x V$, whose members are called the 
\emph{edges} of $G$;
\item
$\LBLv$ is a countable set, whose members are called the 
\emph{vertex labels} of $G$; 
\item
$\LBLe$ is a countable set, whose members are called the 
\emph{edge labels} of $G$; 
\item
$l$ is a partial function from $V \union E$ to $\LBLv \union \LBLe$ 
such that
\begin{itemize}
\item[]
for all $v \in V$ for which $l(v)$ is defined, $l(v) \in \LBLv$ and
\item[]
for all $e \in E$ for which $l(e)$\, is defined, $l(e) \in \LBLe$,
\end{itemize}
called the \emph{labeling function} of $G$;
\item 
$r \in V$, called the \emph{root} of $G$.
\end{itemize}
\end{definition}
The additional graph theoretical notions defined below are also used.
\begin{definition}
Let $G = (V,E,\LBLv,\LBLe,l,r)$ be a rooted labeled directed graph.
Then a \emph{cycle} in $G$ is a sequence 
$v_1\, \ldots\, v_{n+1} \in V^*$
such that, for all $i \in \set{1,\ldots,n}$, $(v_i,v_{i+1}) \in E$,
$\mathrm{card}(\set{v_1,\ldots,v_n}) = n$, and $v_1 = v_{n+1}$.
Let, moreover, $v \in V$.
Then 
the \emph{indegree} of $v$, written $\indeg(v)$, is  
$\mathrm{card}(\set{v' \where (v',v) \in E})$ 
and
the \emph{outdegree} of $v$, written $\outdeg(v)$, is  
$\mathrm{card}(\set{v' \where (v,v') \in E})$. 
\end{definition}

Concerning sequences, we write:
\begin{itemize}
\item
$\cA^\infty$ for the set of all non-empty sequences over the set $\cA$ 
that are finite or countably infinite;
\item
$\emptyseq$ for the empty sequence;
\item
$\seq{a}$ for the sequence having $a$ as sole element;
\item 
$\alpha \conc \alpha'$,  for the concatenation of the sequences 
$\alpha$ and $\alpha'$;
\item
$\alpha[n]$ for the $n$th element of the sequence $\alpha$ if there 
exists a prefix of $\alpha$ of length $n$ and otherwise the last 
element of $\alpha$;
\item
$|\alpha|$ for the length of the sequence $\alpha$.
\end{itemize}

Concerning tuples, we write:
\begin{itemize}
\item
$\boldsymbol{a}(i)$, where $\boldsymbol{a} = (a_1,\ldots,a_n)$, 
for $a_i$;
\item
$\updtup{\boldsymbol{a}}{i}{a}$, 
where $\boldsymbol{a} = (a_1,\ldots,a_n)$, 
for $(a_1,\ldots,a_{i-1},a,a_{i+1},\ldots,a_n)$.
\end{itemize}

Moreover, we write:
\begin{itemize}
\item
$\Natpos$ for the set $\set{n \in \Nat \where n > 0}$ of positive 
natural numbers;
\item
$\cA^n$ for the $n$-fold Cartesian product of the set $\cA$ with itself;
\item
$\cP(\cA)$ for the set of all subsets of the set $\cA$.
\end{itemize}

In Sections~\ref{sect-noninter-algo} and~\ref{sect-inter-algo}, we 
assume the existence of a dummy value $\bot$ that is not a member of 
certain sets.
Concerning $\bot$, we write:
\begin{itemize}
\item
$\cA_\bot$, where $\cA$ is a set such that $\bot \notin \cA$, for 
$\cA \union \set{\bot}$;
\item
$\bot^n$ for the unique member of the set $\set{\bot}^n$.
\end{itemize}

In this paper, we sometimes consider multi-valued functions, 
i.e.\ functions from a set $\cA$ to the set of all subsets of a set 
$\cA'$.
The following notion concerning multi-valued functions is used:
\begin{definition}
Let $\cA$ and $\cA'$ be sets, and
let $f$ and $g$ be functions from $\cA$ to $\cP(\cA')$.
Then $f$ is \emph{smaller then} $g$ if $f(s) \subset g(s)$ for all 
$s \in \cA$.
\end{definition}

Moreover, the following notion of a finitely generated set is used:
\begin{definition}
Let $\cA$ be a set.
Then $\cA$ is a \emph{finitely generated set} if there exist a finite 
subset $\cA'$ of $\cA$ and a finite set $\cF$ of functions on $\cA$ such 
that, for each subset $\cA''$ of $\cA$ that includes $\cA'$, the closure 
of $\cA''$ under the functions in $\cF$ equals $\cA$.
\end{definition}

\section{Informal Notions of Algorithm}
\label{sect-algo-informal}

In this paper, the notion of a classical algorithm concerns the original
notion of an algorithm, i.e.\ the notion that is intuitively 
characterized in standard works from the mathematical and computer 
science literature such as~\cite{Kle67a,Knu97a,Mal70a,Rog67a}.
Seeing the terminology used for proposed generalizations of the original 
notion of an algorithm, a classical algorithm could also be called a 
deterministic non-interactive algorithm.

The characterizations of the notion of a classical algorithm referred to 
above indicate that a classical algorithm is considered to express a 
pattern of behaviour by which all instances of a computational problem 
can be solved.
This calls for a description of a classical computational problem that 
does not refer to the notion of a classical algorithm:
\begin{quote}
A classical computational problem is a problem where, given a value from a 
certain set of input values, a value from a certain set of output values 
that is in a certain relation to the given input value must be produced if 
it exists.
The input values are also called the instances of the problem and an 
output value that is in the certain relation to a given input value is 
also called a solution for the instance concerned.
\end{quote}

From the existing viewpoints on what a classical algorithm is, it 
follows that the following properties must be considered the most 
important ones of a classical algorithm: 
\begin{enumerate}
\item
a classical algorithm is a finite expression of a pattern of behaviour 
by which all instances of a classical computational problem can be 
solved;
\item
the pattern of behaviour expressed by a classical algorithm is made up 
of discrete steps, each of which consists of performing an elementary 
operation or inspecting an elementary condition unless it is the initial 
step or a final step;
\item
the pattern of behaviour expressed by a classical algorithm is such that 
there is one possible step immediately following a step that consists of 
performing an operation;
\item
the pattern of behaviour expressed by a classical algorithm is such that 
there is one possible step immediately following a step that consists of 
inspecting a condition for each outcome of the inspection;
\item
the pattern of behaviour expressed by a classical algorithm is such that 
the initial step consists of inputting an input value of the problem 
concerned;
\item
the pattern of behaviour expressed by a classical algorithm is such 
that, for each input value of the problem concerned for which a correct 
output value exists, a final step is reached after a finite number of 
steps and that final step consists of outputting a correct output value 
for that input value;
\item
the steps involved in the pattern of behaviour expressed by a classical 
algorithm are precisely and unambiguously defined and can be done 
exactly in a finite amount of time.
\end{enumerate}

In the publications in which the viewpoints are expressed from which
properties~1--7 are derived, it is usually mentioned at most in passing 
that a classical algorithm expresses a pattern of behaviour.
Following~\cite{Dij71a}, this point is central in this paper.

By property~3, a classical algorithm is a deterministic algorithm.
I coin the term non-interactive algorithm for the notion of an algorithm 
characterized by the same properties as the notion of a classical 
algorithm except that the restriction to deterministic algorithms is 
dropped, i.e.\ with property~3 replaced by following property:
\begin{enumerate}
\item[3$'$]
the pattern of behaviour expressed by a classical algorithm is such that 
there is at least one possible step immediately following a step that 
consists of performing an operation.
\end{enumerate}

In this paper, properties 1, 2, 3$'$, 4, 5, 6, and 7 are considered to 
give an intuitive characterization of the notion of a non-interactive 
algorithm and will be used as the starting point for the formalization 
of this notion. 

The viewpoints on interactive computation that are expressed in computer 
science publications such as~\cite{Bro15a,GSAS04a,LW01a} are the basis 
of the following description of an interactive computational problem:
\begin{quote}
An interactive computational problem is a problem where, given a 
possibly infinite sequence of values from a certain set of input values, 
a possibly infinite sequence of values from a certain set of output 
values that is in a certain causal relation to the given sequence of 
input values must be produced if it exists. 
Causal here means that, for each $n$, for each two sequences of input 
values with a common prefix of length $n$, for each sequence of output 
values with a prefix of length $n$ that is related to one of those two
sequences of input values, there exists a sequence of output values with
the same prefix that is related to the other of those two sequences of 
input values.
The sequences of input values are called the instances of the problem 
and a sequence of output values that is in the certain relation to a 
given sequence of input values is called a solution for the instance 
concerned.
\end{quote}
Classical computational problems can be considered the simplest 
interactive computational problems.

Classical algorithms are not adequate for solving interactive 
computational problems.
From the viewpoints on interactive computation referred to above, it 
follows that:
\begin{itemize}
\item
with the exception of properties~2, 3, and~6, the most important 
properties of a classical algorithm are also considered to belong to the 
most important properties of an interactive algorithm;
\item
the following generalizations of properties~2 and~6 are considered to 
belong to the most important properties of an interactive algorithm:

\begin{itemize}
\item[2$'$.]
the pattern of behaviour expressed by an interactive algorithm is made 
up of discrete steps, each of which consists of performing an elementary 
operation or inspecting an elementary condition unless it is the initial 
step, a final step, an input step or an output step;
\item[6$'$.]
the pattern of behaviour expressed by an interactive algorithm is such 
that, for each finite sequence of input values of the problem concerned 
for which a correct sequence of output values exists, after this 
sequence has been inputted, a final step is reached after a finite 
number of steps and that final step consists of outputting the last 
output value of a correct sequence of output values;
\end{itemize}
\item
the following property is also considered to belong to the most 
important ones of an interactive algorithm: 
\begin{itemize}
\item[8.]
the pattern of behaviour expressed by an interactive algorithm is such 
that:
\begin{itemize}
\item[a.]
there is one possible step immediately following an output step;
\item[b.] 
an output step is immediately followed by an input step and an input 
step immediately follows an output step; 
\item[c.]
an input step consists of inputting an input value of the problem 
concerned and an output step consists of outputting an output value that
is correct for the sequence of input values inputted so far.
\end{itemize}
\end{itemize}
\end{itemize}

From the viewpoints on interactive computation referred to above, it is
not clear whether property~3 should be considered to belong to the most
important properties of an interactive algorithm.
Therefore, in this paper, properties 1, 2$'$, 3$'$, 4, 5, 6$'$, 7, and 8 
are considered to give an intuitive characterization of the notion of an 
interactive algorithm and will be used as the starting point for the 
formalization of this notion. 

\section{Non-Interactive Algorithms}
\label{sect-noninter-algo}

In this section, the notion of a non-interactive proto-algorithm 
is introduced first.
The thought is that non-interactive algorithms are equivalence classes 
of non-interactive proto-algorithms under some equivalence relation.
That is why three equivalence relations are defined next.
Two of them give bounds between which an appropriate equivalence 
relation must lie.
The third lies in between these two and is likely an appropriate 
equivalence relation.

\subsection{Non-Interactive Proto-Algorithms}
\label{subsect-noninter-algo-proto}

The notion of a non-interactive proto-algorithm will be defined in terms 
of three auxiliary notions.
First, we define these auxiliary notions, starting with the notion of a 
non-interactive alphabet.

\begin{definition}
\label{def-noninter-alphabet}
A \emph{non-interactive alphabet} $\Sigma$ is a couple $(F,P)$, 
where:
\begin{itemize}
\item
$F$ is a countable set of \emph{function symbols} of $\Sigma$;
\item
$P$ is a countable set of \emph{predicate symbols} of $\Sigma$;
\item
$F$ and $P$ are disjoint sets and $\ini,\fin \in F$.
\end{itemize}
\end{definition}
We write $\widehat{F}$ and $\widetilde{F}$, where $F$ is the set of 
function symbols of a non-interactive alphabet, for the sets 
$F \diff \set{\fin}$ and $F \diff \set{\ini,\fin}$, 
respectively.

In the case of a non-interactive alphabet $(F,P)$, the function symbols 
from~$\widetilde{F}$ and predicate symbols from $P$ refer to the 
operations and conditions, respectively, involved in the steps of which 
the pattern of behaviour expressed by a non-interactive algorithm is 
made up.
The function symbols $\ini$ and $\fin$ refer to inputting an input value 
and outputting an output value, respectively.

We are now ready to define the notions of a non-interactive 
$\Sigma$-algorithm graph and a non-interactive $\Sigma$-interpretation.
They concern the pattern of behaviour expressed by a non-interactive 
algorithm.

\begin{definition}
\label{def-noninter-algo-graph}
\sloppy
Let $\Sigma = (F,P)$ be a non-interactive alphabet.
Then a \emph{non-inter\-active $\Sigma$-algorithm graph} $G$ is a rooted 
labeled directed graph $(V,E,\LBLv,\LBLe,l,r)$ such that
\begin{itemize}
\item
$\LBLv = F \union P$;
\item
$\LBLe = \set{0,1}$;
\item
for all $v \in V$:
\begin{itemize}
\item
$\indeg(v) = 0$ iff $v = r$;
\item
$l(v) = \ini$ iff $\indeg(v) = 0$;
\item
$l(v) = \fin$ iff $\outdeg(v) = 0$;
\item
if $l(v) \in F$, then, 
for each $v' \in V$ such that $(v,v') \in E$, $l((v,v'))$ is undefined;
\item
if $l(v)  \in P$, then $\outdeg(v) = 2$ and, 
for the unique $v',v'' \in V$ such that $v' \neq v''$ and 
$(v,v'),(v,v'') \in E$, both $l((v,v'))$ and $l((v,v''))$ are defined 
and $l((v,v')) \neq l((v,v''))$;
\end{itemize}
\item
for all cycles $v_1\, \ldots\, v_{n+1}$ in $G$, 
there exists a $v \in \set{v_1,\ldots,v_n}$ such that 
$l(v) \in F$.
\end{itemize}
\end{definition}
\sloppy
Non-interactive $\Sigma$-algorithm graphs are somewhat reminiscent of 
program schemes as defined, for example, in~\cite{Wey79a}.

In the above definition, the condition on cycles in a non-interactive 
$\Sigma$-algorithm graph excludes infinitely many consecutive steps, 
each of which consists of inspecting a condition.

In the above definition, the conditions regarding the vertices and edges
of a non-interactive $\Sigma$-algorithm graph correspond to the 
essential properties of a non-interactive algorithm described in 
Section~\ref{sect-algo-informal} that concern its structure.
Adding an interpretation of the symbols of the non-interactive alphabet 
$\Sigma$ to a non-interactive $\Sigma$-algorithm graph yields something 
that has all of the essential properties of a non-interactive algorithm 
described in Section~\ref{sect-algo-informal}.

\begin{definition}
\label{def-noninter-interpretation}
Let $\Sigma = (F,P)$ be a non-interactive alphabet.
Then a \emph{non-interactive $\Sigma$-inter\-pretation} $\cI$ is a 
quadruple $(D,\Din,\Dout,I)$, 
where:
\begin{itemize}
\item
$D$ is a set, called the \emph{algorithm domain} of $\cI$;
\item
$\Din$ is a finitely generated set, called the \emph{input domain} of 
$\cI$;
\item
$\Dout$ is a finitely generated set, called the \emph{output domain} of 
$\cI$;
\item
$I$ is a total function from $F \union P$ to the set of all total 
computable functions from $\Din$ to $D$, $D$ to $\Dout$, $D$ to $D$ or 
$D$ to $\set{0,1}$ such that:
\begin{itemize}
\item
$I(\ini)$ is a function from $\Din$ to $D$;
\item
$I(\fin)$ is a function from $D$ to $\Dout$;
\item
for all $f \in \widetilde{F}$, $I(f)$ is a function from $D$ to $D$;
\item
for all $p \in P$, $I(p)$ is a function from $D$ to $\set{0,1}$; 
\end{itemize}
\item
there does not exist a $D' \subset D$ such that:
\begin{itemize}
\item
for all $\din \in \Din$, $I(\ini)(\din) \in D'$;
\item
for all $f \in \widetilde{F}$, for all $d \in D'$, $I(f)(d) \in D'$.
\end{itemize}
\end{itemize}
\end{definition}
The finite generation condition on $\Din$ and $\Dout$ and the minimality 
condition on $D$ in the above definition are considered desirable. 
They guarantee that all elements of $\Din$, $\Dout$, and $D$ have a 
finite representation, which is generally expected of the values 
involved in the steps of an algorithm.
However, these conditions have not yet been shown to be essential in 
establishing results.

The pattern of behavior expressed by a non-interactive algorithm can be 
fully represented by the combination of a non-interactive alphabet 
$\Sigma$, a non-interactive $\Sigma$-algorithm graph $G$, and a 
non-interactive $\Sigma$-interpretation $\cI$.
This brings us to defining the notion of a non-interactive 
proto-algorithm. 

\begin{definition}
\label{def-noninter-algo-proto}
A \emph{non-interactive proto-algorithm} $A$ is a triple 
$(\Sigma,G,\cI)$, where:
\begin{itemize}
\item
$\Sigma$ is a non-interactive alphabet, called the \emph{alphabet} of 
$A$;
\item
$G$ is a non-interactive $\Sigma$-algorithm graph, 
called the \emph{algorithm graph} of $A$;
\item
$\cI$ is a non-interactive $\Sigma$-interpretation, 
called the \emph{interpretation} of $A$.
\end{itemize}
\end{definition}

A distinction can be made between deterministic non-interactive 
proto-algorithms and non-deterministic non-interactive proto-algorithms.
\begin{definition}
\label{def-noninter-algo-proto-det-nondet}
Let $A = (\Sigma,G,\cI)$ be a non-interactive proto-algorithm, where 
$\Sigma = (F,P)$ and $G = (V,E,\LBLv,\LBLe,l,r)$.
Then \emph{$A$ is deterministic} if, for all $v \in V$ with 
$l(v) \in \widehat{F}$, $\outdeg(v) = 1$, and 
\emph{$A$ is non-deterministic} if $A$ is not deterministic.
\end{definition}

Henceforth, we assume a dummy value $\bot$ such that, 
for all non-interactive proto-algorithms $A = (\Sigma,G,\cI)$, where 
$\Sigma = (F,P)$, $G = (V,E,\LBLv,\LBLe,l,r)$, and 
$\cI = (D,\Din,\Dout,I)$, 
$\bot \notin V$, $\bot \notin D$, $\bot \notin \Din$, and
$\bot \notin \Dout$.

The intuition is that a non-interactive proto-algorithm is something 
that goes from one state to another.
\begin{definition} 
\label{def-noninter-algo-state}
Let $A = (\Sigma,G,\cI)$ be a non-interactive proto-algorithm, where 
$\Sigma = (F,P)$, $G = (V,E,\LBLv,\LBLe,l,r)$, and 
$\cI = (D,\Din,\Dout,I)$.
Then a \emph{state} of $A$ is a triple 
$(\din,(v,d),\dout) \in
 {\Din}_\bot \x (V_\bot \x D_\bot) \x {\Dout}_\bot$ such that:
\begin{itemize}
\item
$v = \bot$ iff $d = \bot$;
\item
if $(v,d) = (\bot,\bot)$, then $\din = \bot$ iff $\dout \neq \bot$;
\item
if $(v,d) \neq (\bot,\bot)$, then $\din = \bot$ and $\dout = \bot$.
\end{itemize}
A state $s$ of $A$ is
\begin{itemize}
\item 
an \emph{initial state} of $A$ if
$s \in \Din \x \set{(\bot,\bot)} \x \set{\bot}$;
\item 
a \emph{final state} of $A$ if 
$s \in \set{\bot} \x \set{(\bot,\bot)} \x \Dout$;
\item 
an \emph{internal state} of $A$ if 
$s \in \set{\bot} \x (V \x D) \x \set{\bot}$. 
\end{itemize}
\end{definition}
Henceforth, we write $\cS_A$, where $A$ is a non-interactive 
proto-algorithm, for the set of all states of $A$.
We also write $\Sini_A$, $\Sfin_A$, and $\Sint_A$ for 
the set of all initial states of $\cS_A$, 
the set of all final states of $\cS_A$, and
the set of all internal states of $\cS_A$, respectively.
Moreover, we write $\botc$ for the tuple $(\bot,\bot)$.

An internal state $(\bot,(v,d),\bot)$ of a non-interactive 
proto-algorithm $A$ can be largely explained from the point of view of 
state machines:
\begin{itemize}
\item
$v$ is the control state of $A$;
\item
$d$ is the data state of $A$.
\end{itemize}

Suppose that $A = (\Sigma,G,\cI)$ is a non-interactive proto-algorithm, 
where $\Sigma = (F,P)$, $G = (V,E,\LBLv,\LBLe,l,r)$, and  
$\cI = (D,\Din,\Dout,I)$.
$A$ goes from one state to the next state by making a step, it starts in 
an initial state, and if it does not keep making steps forever, it stops 
in a final state.
The following is an informal explanation of how the state that $A$ is in 
determines what the possible steps to a next state consists of and to 
what next states they lead:
\begin{itemize}
\sloppy
\item
if $A$ is in initial state $(\din,\botc,\bot)$, then a step to a next 
state is possible that consists of applying function $I(\ini)$ to $\din$ 
and that leads to one of the internal states $(\bot,(v',d'),\bot)$ such 
that $(r,v') \in E$ and $I(\ini)(\din) = d'$;
\item
if $A$ is in internal state $(\bot,(v,d),\bot)$ and 
$l(v) \in \widetilde{F}$, then a step to a next state is possible that 
consists of applying function $I(l(v))$ to $d$ and that leads to one of 
the internal states $(\bot,(v',d'),\bot)$ such that $(v,v') \in E$ and 
\mbox{$I(l(v))(d) = d'$};
\item
if $A$ is in internal state $(\bot,(v,d),\bot)$ and $l(v) \in P$, then 
a step to a next state is possible that consists of applying function 
$I(l(v))$ to $d$ and that leads to the unique internal state 
$(\bot,(v',d),\bot)$ such that $(v,v') \in E$ and 
\mbox{$I(l(v))(d) = l((v,v'))$};
\item
if $A$ is in internal state $(\bot,(v,d),\bot)$ and $l(v) = \fin$, then 
a step to a next state is possible that consists of applying function 
$I(\fin)$ to $d$ and that leads to the unique final state 
$(\bot,\botc,\dout)$ such that $I(\fin)(v) = \dout$.
\end{itemize}
This informal explanation of how the state that $A$ is in determines 
what the possible next states are, is formalized by the algorithmic step 
function $\astep_A$ defined in 
Section~\ref{subsect-noninter-step-run-comp}. 

Because non-interactive proto-algorithms are considered too concrete to 
be called non-interactive algorithms, the term non-interactive 
proto-algorithm has been chosen instead of the term non-interactive 
algorithm.
For example, from a mathematical point of view, it is natural to 
consider the behavioral patterns expressed by isomorphic non-interactive 
proto-algorithms to be the same.

It should be intuitively clear what isomorphism of non-interactive 
proto-algo\-rithms is.
For the sake of completeness, here is a precise definition.
\begin{definition}
\label{def-noninter-algo-iso}
\sloppy
Let $A = (\Sigma,G,\cI)$ and $A' = (\Sigma',G',\cI')$ be non-interactive 
proto-algorithms, where $\Sigma = (F,P)$, $\Sigma' = (F',P')$, 
$G = (V,E,\LBLv,\LBLe,l,r)$, $G' = (V',E',\LBLv',\LBLe',l',r')$, 
$\cI = (D,\Din,\Dout,I)$, and $\cI' = (D',\Din',\Dout',I')$.
Then $A$ and $A'$ are \emph{isomorphic}, written $A \iso A'$, 
if there exist bijections 
$\funct{\bijF}{F}{F'}$, $\funct{\bijP}{P}{P'}$, 
$\funct{\bijV}{V}{V'}$, $\funct{\bijD}{D}{D'}$, 
$\funct{\bijI}{\Din}{\Din'}$, $\funct{\bijO}{\Dout}{\Dout'}$, and 
$\funct{\bijB}{\set{0,1}}{\set{0,1}}$ such that:
\begin{itemize}
\item
for all $v,v' \in V$,  
$(v,v') \in E$ iff $(\bijV(v),\bijV(v')) \in E'$;
\item
for all $v \in V$ with $l(v) \in F$,\, $\bijF(l(v)) = l'(\bijV(v))$;
\item
for all $v \in V$ with $l(v) \in P$,\, $\bijP(l(v)) = l'(\bijV(v))$; 
\item
for all $(v,v') \in E$ with $l((v,v'))$ defined, 
$\bijB(l((v,v'))) = l'((\bijV(v),\bijV(v')))$;
\item
$\bijF(\ini) = \ini$ and $\bijF(\fin) = \fin$;
\item
for all $\din \in \Din$, $\bijD(I(\ini)(\din)) = I'(\ini)(\bijI(\din))$;
\item
for all $d \in D$, $\bijO(I(\fin)(d)) = I'(\fin)(\bijD(d))$;
\item
for all $d \in D$ and $f \in \widetilde{F}$, 
$\bijD(I(f)(d)) = I'(\bijF(f))(\bijD(d))$;
\item
for all $d \in D$ and $p \in P$, 
$\bijB(I(p)(d)) = I'(\bijP(p))(\bijD(d))$.
\end{itemize}
\end{definition}

Non-interactive proto-algorithms may also be considered too concrete in 
a way not covered by isomorphism of non-interactive proto-algorithms.
This issue is addressed in Section~\ref{subsect-noninter-algo-equiv} and 
leads there to the introduction of two other equivalence relations.

\sloppy
A non-interactive proto-algorithm could also be defined as a quadruple 
$(D,\Din,\Dout,\overline{G})$ where $\overline{G}$ is a graph that 
differs from a non-interactive $\Sigma$-algorithm graph in that its 
vertex labels are computable functions from $\Din$ to $D$, $D$ to 
$\Dout$, $D$ to $D$ or $D$ to $\set{0,1}$ instead of function and 
predicate symbols from $\Sigma$.
I consider the definition of a non-interactive proto-algorithm given 
earlier more insightful because it isolates as much as possible the 
operations to be performed and the conditions to be inspected from the 
structure of a non-interactive proto-algorithm.

\subsection{On Steps, Runs and What is Computed}
\label{subsect-noninter-step-run-comp}

In Section~\ref{subsect-noninter-algo-proto}, the intuition was given 
that a non-interactive proto-algorithm $A$ is something that goes 
through states.
It was informally explained how the state that it is in determines what 
the possible steps to a next state consists of and to what next states 
they lead.
The algorithmic step function $\astep_A$ that is defined below 
formalizes this.
The computational step function $\cstep_A$ that is also defined below is 
like the algorithmic step function $\astep_A$, but conceals the steps 
that consist of inspecting conditions.

\begin{definition}
\label{def-noninter-algo-step-fnc}
Let $A = (\Sigma,G,\cI)$ be a non-interactive proto-algorithm, 
where $\Sigma = (F,P)$, $G = (V,E,\LBLv,\LBLe,l,r)$, and 
$\cI = (D,\Din,\Dout,I)$.
Then the \emph{algorithmic step function $\astep_A$ induced by $A$} is 
the smallest total function from $\cS_A$ to $\cP(\cS_A) \diff \emptyset$ 
such that for all $d,d' \in D$, $\din \in \Din$, $\dout \in \Dout$, and
$v' \in V$:%
\footnote{Below, we write $S \ni s$ instead of $s \in S$.}
\begin{center}
\renewcommand{\arraystretch}{1}
\begin{tabular}{@{}r@{}l@{\,}l@{}} 
$\astep_A((\din,\botc,\bot))$ & ${} \ni (\bot,(v',d'),\bot)$ \\
\multicolumn{3}{l}{\hspace*{8.3em}
if $l(r) = \ini$, $(r,v') \in E$, and $I(\ini)(\din) = d'$,}
\\
$\astep_A((\bot,(v,d),\bot))$ & ${} \ni (\bot,(v',d'),\bot)$ \\ 
\multicolumn{3}{l}{\hspace*{8.3em}
if $l(v) \in \widetilde{F}$, $(v,v') \in E$, and $I(l(v))(d) = d'$,}
\\
$\astep_A((\bot,(v,d),\bot))$ & ${} \ni (\bot,(v',d),\bot)$ \\
\multicolumn{3}{l}{\hspace*{8.3em}
if $l(v) \in P$, $(v,v') \in E$, and $I(l(v))(d) = l((v,v'))$,}
\\
$\astep_A((\bot,(v,d),\bot))$ & ${} \ni (\bot,\botc,\dout)$ \\ 
\multicolumn{3}{l}{\hspace*{8.3em}
if $l(v) = \fin$ and $I(\fin)(d) = \dout$,}
\\
$\astep_A((\bot,\botc,\dout))$ & ${} \ni (\bot,\botc,\dout)$
\end{tabular}
\end{center}
and the \emph{computational step function $\cstep_A$ induced by $A$} is 
the smallest total function from $\cS_A$ to $\cP(\cS_A) \diff \emptyset$ 
such that, for all $d,d' \in D$, $\din \in \Din$, $\dout \in \Dout$, 
\linebreak[2] $v' \in V$, and $s \in \cS_A$:
\begin{center}
\renewcommand{\arraystretch}{1}
\begin{tabular}{@{}r@{}l@{\,}l@{}} 
$\cstep_A((\din,\botc,\bot))$ & ${} \ni (\bot,(v',d'),\bot)$ \\
\multicolumn{3}{l}{\hspace*{8.3em}
if $l(r) = \ini$, $(r,v') \in E$, and $I(\ini)(\din) = d'$,}
\\
$\cstep_A((\bot,(v,d),\bot))$ & ${} \ni (\bot,(v',d'),\bot)$ \\ 
\multicolumn{3}{l}{\hspace*{8.3em}
if $l(v) \in \widetilde{F}$, $(v,v') \in E$, and $I(l(v))(d) = d'$,}
\\
$\cstep_A((\bot,(v,d),\bot))$ & ${} \ni s$ \\
\multicolumn{3}{l}{\hspace*{8.3em}
if $l(v) \in P$, $(v,v') \in E$, $I(l(v))(d) = l((v,v'))$, and} \\
\multicolumn{3}{l}{\hspace*{8.3em} \phantom{if}
$\cstep_A((\bot,(v',d),\bot)) \ni s$,}
\\
$\cstep_A((\bot,(v,d),\bot))$ & ${} \ni (\bot,\botc,\dout)$ \\ 
\multicolumn{3}{l}{\hspace*{8.3em}
if $l(v) = \fin$ and $I(\fin)(d) = \dout$,}
\\
$\cstep_A((\bot,\botc,\dout))$ & ${} \ni (\bot,\botc,\dout)$.
\end{tabular}
\end{center}
\end{definition}

Below, we define what the possible runs of a non-interactive 
proto-algorithm on an input value are and what is computed by a 
non-interactive proto-algorithm.
To this end, we first give some auxiliary definitions.

The functions $\asruns_A$ and $\csruns_A$ defined below give the sets of 
sequences of states that $A$ goes through from a given state according 
to the algorithmic step function and computational step function, 
respectively.
\begin{definition}
\label{def-noninter-algo-semi-runs-fnc}
Let $A = (\Sigma,G,\cI)$ be a non-interactive proto-algorithm.
Then the \emph{algorithmic semi-run set function $\asruns_A$ induced by 
$A$} and the \emph{computational semi-run set function $\csruns_A$ 
induced by $A$} are the total functions from $\cS_A$ to 
$\cP({\cS_A}^\infty)$ such that for all $s \in \cS_A$:
\begin{center}
\renewcommand{\arraystretch}{1}
\hspace*{1.5em}
\begin{tabular}{@{}l@{}l@{\;}l@{}} 
$\asruns_A(s)$ & ${} = 
 \set{\seq{s} \conc \sigma \where
      \Lexists{s' \in \astep_A(s)}{\sigma \in \asruns_A(s')}}$ &
if $s \notin \Sfin_A$,
\\
$\asruns_A(s)$ & ${} = \set{\seq{s}}$ & 
if $s \in \Sfin_A$.
\end{tabular}
\end{center} 
and
\begin{center}
\renewcommand{\arraystretch}{1}
\hspace*{1.5em}
\begin{tabular}{@{}l@{}l@{\;}l@{}} 
$\csruns_A(s)$ & ${} = 
 \set{\seq{s} \conc \sigma \where
      \Lexists{s' \in \cstep_A(s)}{\sigma \in \csruns_A(s')}}$ &
if $s \notin \Sfin_A$,
\\
$\csruns_A(s)$ & ${} = \set{\seq{s}}$ & 
if $s \in \Sfin_A$.
\end{tabular}
\end{center} 
A $\sigma \in {\cS_A}^\infty$ is called an \emph{algorithmic semi-run} 
if there exists an $s \in \cS_A$ such that $\sigma \in \asruns_A(s)$. 
\end{definition}
Henceforth, we write $\Asruns_A$, where $A$ is a non-interactive 
proto-algorithm, for the set of all algorithmic semi-runs.

When defining what is computed by a non-interactive proto-algorithm, a 
distinction must be made between convergent algorithmic semi-runs and 
divergent semi-runs.
\begin{definition}
\label{def-noninter-algo-div-conv}
Let $A = (\Sigma,G,\cI)$ be a non-interactive proto-algorithm, and 
let $\sigma \in \Asruns_A$. 
Then \emph{$\sigma$ is divergent} if there exists a suffix $\sigma'$ of 
$\sigma$ such that $\sigma' \in {\Sint_A}^\infty$, and 
\emph{$\sigma$ is convergent} if $\sigma$ is not divergent.
\end{definition}

The function $\outputx_A$ defined below gives the output value 
eventually produced by a given convergent algorithmic semi-run.
\begin{definition}
\label{def-noninter-algo-output-val-fnc}
Let $A = (\Sigma,G,\cI)$ be a non-interactive proto-algorithm.
Then the \emph{output value extraction function 
$\outputx_A$ for $A$} is the total function from 
$\set{\sigma \in \Asruns_A \where \sigma \mathrm{\;is\;convergent}}$ to 
${\Dout}$ such that for all $(\din,c,\dout) \in \cS_A$ and 
$\stseq \in \cS_A^\infty \union \set{\emptyseq}$ such that 
$\seq{(\din,c,\dout)} \conc \stseq \in \Asruns_A$ and
$\seq{(\din,c,\dout)} \conc \stseq$ is convergent: 

\begin{center}
\renewcommand{\arraystretch}{1}
\hspace*{1.5em}
\begin{tabular}{@{}l@{}l@{\;}l@{}} 
$\outputx_A(\seq{(\din,c,\dout)} \conc \sigma)$ & 
${} = \dout$ &              if $\dout \neq \bot$, 
\\
$\outputx_A(\seq{(\din,c,\dout)} \conc \sigma)$ & 
${} = \outputx_A(\sigma)$ & if $\dout = \bot$.
\end{tabular}
\end{center} 
\end{definition}

A run of a non-interactive proto-algorithm is a semi-run of the 
non-interactive proto-algorithm concerned that starts in an initial 
state.
The following definition concerns what the possible runs of a 
non-interactive proto-algorithm on an input value are.
As with steps, a distinction is made between algorithmic runs and 
computational runs.
\begin{definition}
\label{def-noninter-algo-runs}
Let $A = (\Sigma,G,\cI)$ be a non-interactive proto-algorithm, where
$\cI = (D,\Din,\Dout,I)$, and
let $\din \in \Din$.
Then the \emph{algorithmic run set of $A$ on $\din$}, 
written $\arun_A(\din)$, is $\asruns_A((\din,\botc,\bot))$ and
the \emph{computational run set of $A$ on $\din$}, 
written $\crun_A(\din)$, is $\csruns_A((\din,\botc,\bot))$.
\end{definition}

The following definition concerns what is computed by a non-interactive 
proto-algorithm.
\begin{definition}
\label{def-noninter-algo-computed-rel}
Let $A = (\Sigma,G,\cI)$ be a non-interactive proto-algorithm, where 
$\cI = (D,\Din,\Dout,I)$. 
Then the \emph{relation $\widehat{A}$ computed by $A$} is the relation 
from $\Din$ to $\Dout$ such that 
for all $\din \in \Din$ and $\dout \in \Dout$:
\begin{center}
\renewcommand{\arraystretch}{1}
\begin{tabular}{@{}l@{}} 
$(\din,\dout) \in \widehat{A}$ iff 
there exists a convergent $\sigma \in \arun_A(\din)$ such that 
$\dout = \outputx_A(\sigma)$.
\end{tabular}
\end{center}
\end{definition}
If $A$ is a deterministic non-interactive proto-algorithm, then the 
relation $\widehat{A}$ computed by $A$ is functional, i.e.\ 
$\widehat{A}$ is (the graph of) a partial function.

\subsection{Algorithmic and Computational Equivalence}
\label{subsect-noninter-algo-equiv}

If a non-interactive proto-algorithm $A'$ can mimic a non-interactive 
proto-algorithm $A$ step-by-step, then we say that $A$ is 
algorithmically simulated by $A'$.
If the steps that consist of inspecting conditions are ignored, then we 
say that $A$ is computationally simulated by $A'$.
Algorithmic and computational simulation can be formally defined using 
the step functions defined in 
Section~\ref{subsect-noninter-step-run-comp}.

\begin{definition}
\label{def-noninter-algo-sim}
Let $A = (\Sigma,G,\cI)$ and $A' = (\Sigma',G',\cI')$ be non-interactive 
proto-algorithms.
Then an \emph{algorithmic simulation of $A$ by $A'$} is a set
$R \subseteq \cS_A \x \cS_{A'}$ such that: 
\begin{itemize}
\item
for all $s \in \cS_A$ and $s' \in \cS_{A'}$:
\begin{itemize}
\item
if $\sini \in \Sini_A$, then there exists an 
$\sini' \in \Sini_{A'}$ such that $(\sini,\sini') \in R$;
\item
if $\sfin' \in \Sfin_{A'}$, then there exists an  
$\sfin \in \Sfin_A$ such that $(\sfin,\sfin') \in R$;
\item
if $(s,s') \in R$ and $t \in \astep_A(s)$, 
then there exists a $t' \in \astep_{A'}(s')$ such that 
{$(t,t') \in R$};
\end{itemize}
\item
for all $(s,s') \in R$:
\begin{itemize}
\item
$s \in \Sini_A$ iff $s' = \Sini_{A'}$;
\item 
$s \in \Sfin_A$ iff $s' = \Sfin_{A'}$
\end{itemize}
\end{itemize}
and a \emph{computational simulation of $A$ by $A'$} is a set
$R \subseteq \cS_A \x \cS_{A'}$ such that: 
\begin{itemize}
\item
for all $s \in \cS_A$ and $s' \in \cS_{A'}$:
\begin{itemize}
\item
if $\sini \in \Sini_A$, then there exists an 
$\sini' \in \Sini_{A'}$ such that $(\sini,\sini') \in R$;
\item
if $\sfin' \in \Sfin_{A'}$, then there exists an  
$\sfin \in \Sfin_A$ such that $(\sfin,\sfin') \in R$;
\item
if $(s,s') \in R$ and $t \in \cstep_A(s)$, 
then there exists a $t' \in \cstep_{A'}(s')$ such that 
{$(t,t') \in R$};
\end{itemize}
\item
for all $(s,s') \in R$:
\begin{itemize}
\item
$s \in \Sini_A$ iff $s' = \Sini_{A'}$;
\item 
$s \in \Sfin_A$ iff $s' = \Sfin_{A'}$.
\end{itemize}
\end{itemize}
$A$ \emph{is algorithmically simulated by} $A'$, written $A \asim A'$,
if there exists an algorithmic simulation $R$ of $A$ by $A'$. 
\\ 
$A$ \emph{is computationally simulated by} $A'$, written $A \csim A'$, 
if there exists a computational simulation $R$ of $A$ by $A'$.
\\ 
$A$ \emph{is algorithmically equivalent to} $A'$, written $A \aeqv A'$, 
if there exist an algorithmic simulation $R$ of $A$ by $A'$ and an 
algorithmic simulation $R'$ of $A'$ by $A$ such that $R' = R^{-1}$.
\\
$A$ \emph{is computationally equivalent to} $A'$, written $A \ceqv A'$, 
if there exist a computational simulation $R$ of $A$ by $A'$ and a 
computational simulation $R'$ of $A'$ by $A$ such that $R' = R^{-1}$.
\end{definition}

The conditions imposed on an algorithmic or computational simulation $R$ 
of a non-interactive proto-algorithm $A$ by a non-interactive 
proto-algorithm $A'$ include, in addition to the usual transfer 
conditions, also conditions that guarantee that a state of $A$ is only 
related by $R$ to a state of $A'$ of the same kind (initial, final or 
internal).
Lemma~\ref{lemma-noninter-simulation} given below, used in the proof of 
the theorem that follows it, would not hold without the additional 
conditions.

There may be states of a non-interactive proto-algorithm in which there 
is a choice from multiple possible steps to a next state.
The condition $R' = R^{-1}$ imposed on simulations $R$ and $R'$ 
witnessing (algorithmic or computational) equivalence of non-interactive 
proto-algorithms $A$ and $A'$ guarantees that $A$ and $A'$ have the same 
choice structure.

\begin{lemma}
\label{lemma-noninter-simulation}
Let $A = (\Sigma,G,\cI)$ and $A' = (\Sigma',G',\cI')$ be 
non-interactive proto-algorithms, 
where $\Sigma = (F,P)$, $\Sigma' = (F',P')$, 
$G = (V,E,\LBLv,\LBLe,l,r)$, $G' = (V',E',\LBLv',\LBLe',l',r')$, 
$\cI = (D,\Din,\Dout,I)$, and $\cI' = (D',\Din',\Dout',I')$, and
let $R \subseteq \cS_A \x \cS_{A'}$ and 
$\funct{\fncI}{\Din}{\Din'}$ be such that, for all $\din \in \Din$, 
$((\din,\botc,\bot),(\fncI(\din),\botc,\bot)) \in R$.
Then $R$ is an algorithmic simulation of $A$ by $A'$ only if,
for all $\din \in \Din$, for all $\sigma \in \arun_A(\din)$,
there exists a $\sigma' \in \arun_{A'}(\fncI(\din))$ such that,
for all $n \in \Natpos$, $(\sigma[n],\sigma'[n]) \in R$. 
\end{lemma}
\begin{proof}
Let $R$ be an algorithmic simulation of $A$ by $A'$ and 
$\funct{\fncI}{\Din}{\Din'}$ be such that, for all $\din \in \Din$, 
$((\din,\botc,\bot),(\fncI(\din),\botc,\bot)) \in R$, 
let $\din \in \Din$, and let $\sigma \in \arun_A(\din)$.
Then we can easily construct a $\sigma' \in \arun_{A'}(\fncI(\din))$ 
such that, for all $n \in \Natpos$, $(\sigma[n],\sigma'[n]) \in R$,
using the conditions imposed on algorithmic simulations. 
\qed
\end{proof}

The following theorem tells us that, 
if a non-interactive proto-algorithm $A$ is algorithmically simulated by 
a non-interactive proto-algorithm $A'$, then 
(a)~the relation computed by $A'$ models the relation computed by $A$ 
(in the sense of e.g.~\cite{Jon90a}) and
(b)~for each convergent algorithmic run of $A$, the simulation results
in a convergent algorithmic run of $A'$ consisting of the same number 
of algorithmic steps.
\begin{theorem}
\label{theorem-noninter-alg-equiv}
Let $A = (\Sigma,G,\cI)$ and $A' = (\Sigma',G',\cI')$ be 
non-interactive proto-algorithms, where 
$\cI = (D,\Din,\Dout,I)$, and $\cI' = (D',\Din',\Dout',I')$.
Then $A \asim A'$ only if there exist total functions 
$\funct{\fncI}{\Din}{\Din'}$ and $\funct{\fncO}{\Dout'}{\Dout}$ 
such that: 
\begin{enumerate}
\item[(1)]
for all $\din \in \Din$ and $\dout \in \Dout$,
$(\din,\dout) \in \widehat{A}$ only if 
there exists a $\dout' \in \Dout'$ such that
$(\fncI(\din),\dout') \in \widehat{A'}$ and $\fncO(\dout') =  \dout$;
\item[(2)]
for all $\din \in \Din$, 
for all convergent $\sigma \in \arun_A(\din)$, 
there exists a convergent $\sigma' \in \arun_{A'}(\fncI(\din))$ with
$\fncO(\outputx_{A'}(\sigma')) = \outputx_A(\sigma)$ such that 
$|\sigma| = |\sigma'|$.
\end{enumerate}
\end{theorem}
\begin{proof}
Assume that $A \asim A'$.
Let $R$ be an algorithmic simulation of $A$ by $A'$,
let $\fncI$ be a function from $\Din$ to $\Din'$ such that, 
for all $\din \in \Din$, 
$((\din,\botc,\bot),(\fncI(\din),\botc,\bot)) \in R$, and
let $\fncO$ be a function from $\Dout'$ to $\Dout$ such that, 
for all $\dout' \in \Dout'$, 
$((\bot,\botc,\fncO(\dout')),(\bot,\botc,\dout')) \in R$.
The algorithmic simulation $R$ exists by the definition of $\asim$, and
the functions $\fncI$ and $\fncO$ exist by the definition of an 
algorithmic simulation.
By Lemma~\ref{lemma-noninter-simulation}, 
for all $\din \in \Din$, for all $\sigma \in \arun_A(\din)$,
there exists a $\sigma' \in \arun_{A'}(\fncI(\din))$ such that,
for all $n \in \Natpos$, $(\sigma[n],\sigma'[n]) \in R$.

Let $\din \in \Din$, and
let $\sigma \in \arun_A(\din)$ and $\sigma' \in \arun_{A'}(\fncI(\din))$ 
be such that, for all $n \in \Natpos$, $(\sigma[n],\sigma'[n]) \in R$.
Then from the definition of an algorithmic simulation, it immediately 
follows that, for all $n \in \Natpos$:
\begin{enumerate} 
\item[(a)]
$\sigma[n] \notin \Sint_A$ only if $\sigma'[n] \notin \Sint_{A'}$; 
\item[(b)]
for all $\din \in \Din$ and $\dout \in \Dout$:
\begin{itemize}
\item
$\sigma[n] = (\din,\botc,\bot)$ only if 
$\sigma'[n] = (\fncI(\din),\botc,\bot)$;
\item
$\sigma[n] = (\bot,\botc,\dout)$ only if 
there exists a $\dout' \in \Dout'$ such that 
$\sigma'[n] = (\bot,\botc,\dout')$ and $\fncO(\dout') = \dout$.
\end{itemize}
\end{enumerate}
By the definition of the relation computed by a non-interactive 
proto-algorithm, both (1) and~(2) follow immediately from~(a) and~(b).
\qed
\end{proof}
It is easy to see that Theorem~\ref{theorem-noninter-alg-equiv} goes 
through as far as (1) is concerned if algorithmic simulation is replaced 
by computational simulation.
However, (2) does not go through if algorithmic simulation is replaced by 
computational simulation.

The following theorem tells us how isomorphism, algorithmic equivalence, 
and computational equivalence are related.
\begin{theorem}
\label{theorem-noninter-equivs}
Let $A$ and $A'$ be non-interactive proto-algorithms.
Then: 
\begin{trivlist}
\item[]
$\qquad$ (1) $\;$ $A \iso A'$ only if $A \aeqv A'$ 
$\qquad$ (2) $\;$ $A \aeqv A'$ only if $A \ceqv A'$.
\end{trivlist}
\end{theorem}
\begin{proof}
Let $A = (\Sigma,G,\cI)$ and $A' = (\Sigma',G',\cI')$ be non-interactive 
proto-algorithms, where $\Sigma = (F,P)$, $\Sigma' = (F',P')$, 
$G = (V,E,\LBLv,\LBLe,l,r)$, $G' = (V',E',\LBLv',\LBLe',l',r')$, 
$\cI = (D,\Din,\Dout,I)$, and $\cI' = (D',\Din',\Dout',I')$.

Part~1.
Assume that $A \iso A'$.
Let $\bijV$, $\bijD$, $\bijI$, and $\bijO$ be bijections as in the 
definition of $\iso$, 
let $\bijVb$, $\bijDb$, $\bijIb$, and $\bijOb$ be the extensions of 
$\bijV$, $\bijD$, $\bijI$, and $\bijO$, respectively, with the dummy 
value $\bot$ such that $\bijVb(\bot) = \bot$, $\bijDb(\bot) = \bot$, 
$\bijIb(\bot) = \bot$, and $\bijOb(\bot) = \bot$, and
let $\beta$ be the bijection from $\cS_A$ to $\cS_{A'}$ defined by 
$\beta(\din,(v,d),\dout) = 
 (\bijIb(\din), (\bijVb(v), \bijDb(d)), \bijOb(\dout))$. 
Moreover, let $R = \set{(s,\beta(s)) \where s \in \cS_A}$ and
 $R' = \set{(s,\beta^{-1}(s)) \where s \in \cS_{A'}}$. 
Then $R$ is an algorithmic simulation of $A$ by $A'$ and $R'$ is an 
algorithmic simulation of $A'$ by $A$.
This is easily proved by showing that the conditions from the definition 
of an algorithmic simulation are satisfied for all $(s,s') \in R$ and
for all $(s,s') \in R'$, respectively.
Moreover, it follows immediately from the definitions of $R$ and $R'$ 
that $R' = R^{-1}$.
Hence, $A \aeqv A'$.

Part~2.
Assume that $A \aeqv A'$.
Let $R$ be an algorithmic simulation of $A$ by $A'$ such that $R^{-1}$ 
is an algorithmic simulation of $A'$ by $A$.
Then $R$ is also a computational simulation of $A$ by $A'$ and $R^{-1}$ 
is also an computational simulation of $A'$ by $A$.
This is easily proved by showing that the conditions from the definition 
of a computational simulation are satisfied for all $(s,s') \in R$ and
for all $(s,s') \in R'$, respectively.
Hence, $A \ceqv A'$.
\qed
\end{proof}
The opposite implications do not hold in general.
That is, there exist non-interactive proto-algorithms $A$ and $A'$ for 
which it does not hold that $A \iso A'$ if $A \aeqv A'$ and there exist 
non-interactive proto-algorithms $A$ and $A'$ for which it does not hold 
that $A \aeqv A'$ if $A \ceqv A'$.
In both cases, the construction of a general illustrating example is 
described in~\cite{Mid24a}.

The definition of algorithmic equivalence suggests that it is reasonable
to consider the patterns of behaviour expressed by algorithmically 
equivalent non-interactive proto-algorithms the same.
This suggests in turn that non-interactive algorithms can be considered 
equivalence classes of non-interactive proto-algorithms under 
algorithmic equivalence.
The definition of computational equivalence does not suggest that it is 
reasonable to consider the patterns of behaviour expressed by 
computationally equivalent non-interactive proto-algorithms the same 
because steps that consist of inspecting a condition are treated as if 
they do not belong to the patterns of behaviour.
The relevance of the computational equivalence relation is that any 
equivalence relation that captures the sameness of the patterns of 
behaviour expressed by non-interactive proto-algorithms to a higher 
degree than the algorithmic equivalence relation must be finer than the 
computational equivalence relation.

\section{Interactive Algorithms}
\label{sect-inter-algo}

In this section, the notion of an interactive proto-algorithm is 
introduced first.
As with non-interactive algorithms, the thought is that interactive 
algorithms are equivalence classes of interactive proto-algorithms under 
some equivalence relation.
Again, three equivalence relations are defined next, two of which give 
bounds between which an appropriate equivalence relation must lie and 
one of which is likely an appropriate equivalence relation.

\subsection{Interactive Proto-Algorithms}
\label{subsect-inter-algo-proto}

Like the notion of a non-interactive proto-algorithm, the notion of an 
interactive proto-algorithm will be defined in terms of three auxiliary 
notions. 
First, we define these auxiliary notions, starting with the notion of an 
interactive alphabet.
Whereas there are two special function symbols in a non-interactive 
alphabet, there are four special function symbols in an interactive 
alphabet.

\begin{definition}
\label{def-inter-alphabet}
An \emph{interactive alphabet} $\Sigma$ is a couple $(F,P)$, where:
\begin{itemize}
\item
$F$ is a countable set of \emph{function symbols} of $\Sigma$;
\item
$P$ is a countable set of \emph{predicate symbols} of $\Sigma$;
\item
$F$ and $P$ are disjoint sets and $\ini,\fin,\inp,\outp \in F$.
\end{itemize}
\end{definition}
We write $\widehat{F}$ and $\widetilde{F}$, where $F$ is the set of 
function symbols of an interactive alphabet, for the sets 
$F \diff \set{\fin}$ and $F \diff \set{\ini,\fin,\inp,\outp}$, 
respectively.

In the case of an interactive alphabet $(F,P)$, the function symbols 
from $\widetilde{F}$ and predicate symbols from $P$ refer to the 
operations and conditions, respectively, involved in the steps of which 
the pattern of behaviour expressed by an interactive algorithm is made 
up.
The function symbols $\ini$ and $\fin$ refer to inputting a first input 
value and outputting a last output value, respectively, and
the function symbols $\inp$ and $\outp$ refer to inputting a non-first 
input value and outputting a non-last output value, respectively.

We are now ready to define the notions of an interactive 
$\Sigma$-algorithm graph and an interactive $\Sigma$-interpretation.
They concern the pattern of behaviour expressed by an interactive
algorithm.

The definition of the notion of an interactive $\Sigma$-algorithm graph 
differs from the definition of the notion of a non-interactive 
$\Sigma$-algorithm graph only by the addition of two conditions 
concerning the vertex label $\outp$.

\begin{definition}
\label{def-inter-algo-graph}
Let $\Sigma = (F,P)$ be an interactive alphabet.
Then an \emph{interactive $\Sigma$-algorithm graph} $G$ is a rooted 
labeled directed graph $(V,E,\LBLv,\LBLe,l,r)$ such that
\begin{itemize}
\item
$\LBLv = F \union P$;
\item
$\LBLe = \set{0,1}$;
\item
for all $v \in V$:
\begin{itemize}
\item
$\indeg(v) = 0$ iff $v = r$;
\item
$l(v) = \ini$ iff $\indeg(v) = 0$;
\item
$l(v) = \fin$ iff $\outdeg(v) = 0$;
\item
$l(v) = \outp$ only if $\outdeg(v) = 1$;
\item
if $l(v) \in F$, then, 
for each $v' \in V$ such that $(v,v') \in E$, $l((v,v'))$ is undefined;
\item
if $l(v)  \in P$, then $\outdeg(v) = 2$ and, 
for the unique $v',v'' \in V$ such that $v' \neq v''$ and 
$(v,v'),(v,v'') \in E$, both $l((v,v'))$ and $l((v,v''))$ are defined 
and $l((v,v')) \neq l((v,v''))$;
\end{itemize}
\item
for all $(v,v') \in E$, $l(v) = \outp$ iff $l(v') = \inp$;
\item
for all cycles $v_1\, \ldots\, v_{n+1}$ in $G$, 
there exists a $v \in \set{v_1,\ldots,v_n}$ such that $l(v) \in F$.
\end{itemize}
\end{definition}

In the above definition, the conditions regarding the vertices and edges
of an interactive $\Sigma$-algorithm graph correspond to the essential 
properties of an interactive algorithm mentioned in 
Section~\ref{sect-algo-informal} that concern its structure.
Adding an interpretation of the symbols of the interactive alphabet 
$\Sigma$ to an interactive $\Sigma$-algorithm graph yields something 
that has all of the mentioned essential properties of an interactive 
algorithm described in Section~\ref{sect-algo-informal}.

The definition of the notion of an interactive $\Sigma$-interpretation 
differs from the definition of the notion of a non-interactive 
$\Sigma$-interpretation only by the addition of interpretation 
conditions of the special function symbols $\inp$ and $\outp$ and the 
adaptation of the minimality condition.

\begin{definition}
\label{def-inter-interpretation}
Let $\Sigma = (F,P)$ be an interactive alphabet.
Then a \emph{interactive $\Sigma$-interpretation} $\cI$ is a quadruple 
$(D,\Din,\Dout,I)$, 
where:
\begin{itemize}
\item
$D$ is a set, called the \emph{algorithm domain} of 
$\cI$;
\item
$\Din$ is a finitely generated set, called the \emph{input domain} of $\cI$;
\item
$\Dout$ is a finitely generated set, called the \emph{output domain} of $\cI$;
\item
$I$ is a total function from $F \union P$ to the set of all total 
computable functions from $\Din$ to $D$, $D$ to $\Dout$, 
$D \x \Din$ to $D$, $D$ to $D$ or $D$ to $\set{0,1}$ such that:
\begin{itemize}
\item
$I(\ini)$ is a function from $\Din$ to $D$;
\item
$I(\fin)$ is a function from $D$ to $\Dout$;
\item
$I(\inp)$ is a function from $D \x \Din$ to $D$;
\item
$I(\outp)$ is a function from $D$ to $\Dout$;
\item
for all $f \in \widetilde{F}$, $I(f)$ is a function from $D$ 
to $D$;
\item
for all $p \in P$, $I(p)$ is a function from $D$ to 
$\set{0,1}$;
\end{itemize}
\item
there does not exist a $D' \subset D$ such that:
\begin{itemize}
\item
for all $\din \in \Din$, $I(\ini)(\din) \in D'$;
\item
for all $\din \in \Din$, for all $d \in D'$, $I(\inp)(d,\din) \in D'$;
\item
for all $f \in \widetilde{F}$, for all $d \in D'$, $I(f)(d) \in D'$.
\end{itemize}
\end{itemize}
\end{definition}
As with non-interactive interpretations, the finite generation condition 
on $\Din$ and $\Dout$ and the minimality condition on $D$ in the above 
definition are considered desirable. 
Again, these conditions have not yet been shown to be essential in 
establishing results.

The pattern of behavior expressed by an interactive algorithm can be 
fully represented by the combination of an interactive alphabet 
$\Sigma$, an interactive $\Sigma$-algorithm graph $G$, and an 
interactive $\Sigma$-interpretation $\cI$.
This brings us to defining the notion of an interactive proto-algorithm. 

\begin{definition}
\label{def-inter-algo-proto}
An \emph{interactive proto-algorithm} $A$ is a triple $(\Sigma,G,\cI)$, 
where:
\begin{itemize}
\item
$\Sigma$ is an interactive alphabet, called the \emph{alphabet} of $A$;
\item
$G$ is an interactive $\Sigma$-algorithm graph, 
called the \emph{algorithm graph} of $A$;
\item
$\cI$ is an interactive $\Sigma$-interpretation, 
called the \emph{interpretation} of $A$.
\end{itemize}
\end{definition}

As with non-interactive proto-algorithms, a distinction can be made 
between deterministic interactive proto-algorithms and non-deterministic 
interactive proto-algorithms.
\begin{definition}
\label{def-inter-algo-proto-det-nondet}
Let $A = (\Sigma,G,\cI)$ be an interactive proto-algorithm, where 
$\Sigma = (F,P)$ and $G = (V,E,\LBLv,\LBLe,l,r)$.
Then \emph{$A$ is deterministic} if, for all $v \in V$ with 
$l(v) \in \widehat{F}$, $\outdeg(v) = 1$, and 
\emph{$A$ is non-deterministic} if $A$ is not deterministic.
\end{definition}

As with non-interactive proto-algorithms, we assume a dummy value $\bot$ 
such that, for all interactive proto-algorithms $A = (\Sigma,G,\cI)$, 
where $\Sigma = (F,P)$, $G = (V,E,\LBLv,\LBLe,l,r)$, and 
$\cI = (D,\Din,\Dout,I)$, $\bot \notin V$, $\bot \notin D$, 
$\bot \notin \Din$, and $\bot \notin \Dout$.

The intuition is that an interactive proto-algorithm, like a 
non-interactive proto-algorithm, is something that goes from one state 
to another.
Whereas non-interactive proto-algorithms have three kinds of states, 
interactive proto-algorithm have four kinds of states. 
\begin{definition}
\label{def-inter-algo-state}
Let $A = (\Sigma,G,\cI)$ be an interactive proto-algorithm, where 
$\Sigma = (F,P)$, $G = (V,E,\LBLv,\LBLe,l,r)$, and 
$\cI = (D,\Din,\Dout,I)$.
Then a \emph{state} of $A$ is a triple 
$(\din,(v,d),\dout) \in
 {\Din}_\bot \x (V_\bot \x D_\bot) \x {\Dout}_\bot$ such that:
\begin{itemize}
\item
$v = \bot$ iff $d = \bot$;
\item
if $(v,d) = (\bot,\bot)$, then $\din = \bot$ iff $\dout \neq \bot$;
\item
if $(v,d) \neq (\bot,\bot)$, then $\din = \bot$ iff $\dout = \bot$;
\item
if $\din = \bot$ and $\dout = \bot$, then $l(v) \neq \inp$;
\item
if $\din \neq \bot$ and $\dout \neq \bot$, then $l(v) = \inp$.
\end{itemize}
A state $s$ of $A$ is
\begin{itemize}
\item 
an \emph{initial state} of $A$ if
$s \in \Din \x \set{(\bot,\bot)} \x \set{\bot}$;
\item 
a \emph{final state} of $A$ if 
$s \in \set{\bot} \x \set{(\bot,\bot)} \x \Dout$;
\item 
an \emph{internal state} of $A$ if 
$s \in \set{\bot} \x (V \x D) \x \set{\bot}$;
\item
an \emph{interaction state} of $A$ if 
$s \in \Din \x (V \x D) \x \Dout$. 
\end{itemize}
\end{definition}
As with non-interactive proto-algorithms, we write $\cS_A$, where $A$ is 
an interactive proto-algorithm, for the set of all states of $A$.
We also still write $\Sini_A$, $\Sfin_A$, and $\Sint_A$ for 
the set of all initial states of $\cS_A$, 
the set of all final states of $\cS_A$, and
the set of all internal states of $\cS_A$, respectively.
Moreover, we still write $\botc$ for the tuple $(\bot,\bot)$.

Suppose that $A = (\Sigma,G,\cI)$ is an interactive proto-algorithm, 
where $\Sigma = (F,P)$, $G = (V,E,\LBLv,\LBLe,l,r)$, and 
$\cI = (D,\Din,\Dout,I)$.
$A$ goes from one state to the next state by making a step, it starts in 
an initial state, and if it does not keep making steps forever, it stops 
in a final state.
The informal explanation of how the state that $A$ is in determines what 
the possible steps to a next state consists of and to what next states 
they lead, given in Section~\ref{subsect-noninter-algo-proto} for 
non-interactive proto-algorithms applies also to interactive 
proto-algorithms, but does not cover internal states $(\bot,(v,d),\bot)$ 
with $l(v) = \outp$ and interaction states.
The following additional explanation is about these states:
\begin{itemize}
\item
if $A$ is in internal state $(\bot,(v,d),\bot)$ with 
$l(v) = \outp$, then a step to a next state is possible that consists of 
applying function $I(\outp)$ to $d$ and that leads to one of the 
interaction states $(\din,(v',d),\dout)$ such that $(v,v') \in E$, 
$I(\outp)(d) = \dout$, and $\din \in \Din$;
\item
if $A$ is in interaction state $(\din,(v,d),\dout)$, then a step to a 
next state is possible that consists of applying function $I(\inp)$ to 
$(d,\din)$ and that leads to one of the internal states 
$(\bot,(v',d'),\bot)$ such that $(v,v') \in E$ and 
$I(\inp)(d,\din) = d'$.
\end{itemize}
This informal explanation of how the state that $A$ is in determines 
what the possible next states are, is formalized by the algorithmic step 
function $\astep_A$ defined in Section~\ref{subsect-inter-step-run-comp}. 

Similar to non-interactive proto-algorithms, interactive 
proto-algorithms are considered too concrete to be called interactive 
algorithms.
For example, it is natural to consider the behavioral patterns expressed 
by isomorphic interactive proto-algorithms to be the same.

As with non-interactive algorithms, it should be intuitively clear what 
isomorphism of interactive proto-algorithms is.
For the sake of completeness, a precise definition is also provided 
here.
\begin{definition}
\label{def-inter-algo-iso}
Let $A = (\Sigma,G,\cI)$ and $A' = (\Sigma',G',\cI')$ be interactive
proto-al\-gorithms, where
$\Sigma = (F,P)$, $\Sigma' = (F',P')$, 
$G = (V,E,\LBLv,\LBLe,l,r)$, $G' = (V',E',\LBLv',\LBLe',l',r')$, 
$\cI = (D,\Din,\Dout,I)$, and $\cI' = (D',\Din',\Dout',I')$.
Then $A$ and $A'$ are \emph{isomorphic}, written $A \iso A'$, 
if there exist bijections 
$\funct{\bijF}{F}{F'}$, $\funct{\bijP}{P}{P'}$, $\funct{\bijV}{V}{V'}$, 
$\funct{\bijD}{D}{D'}$, $\funct{\bijI}{\Din}{\Din'}$,
$\funct{\bijO}{\Dout}{\Dout'}$, and 
$\funct{\bijB}{\set{0,1}}{\set{0,1}}$ such that:
\begin{itemize}
\item
for all $v,v' \in V$, 
$(v,v') \in E$ iff $(\bijV(v),\bijV(v')) \in E'$;
\item
for all $v \in V$ with $l(v) \in F$, $\bijF(l(v)) = l'(\bijV(v))$;
\item
for all $v \in V$ with $l(v) \in P$, $\bijP(l(v)) = l'(\bijV(v))$; 
\item
for all $(v,v') \in E$ with $l((v,v'))$ is defined, 
$\bijB(l((v,v'))) = l'((\bijV(v),\bijV(v')))$;
\item
$\bijF(\ini) = \ini$ and $\bijF(\fin) = \fin$;
\item
for all $\din \in \Din$, $\bijD(I(\ini)(\din)) = I'(\ini)(\bijI(\din))$;
\item
for all $d \in D$, $\bijO(I(\fin)(d)) = I'(\fin)(\bijD(d))$;
\item
for all $d \in D$, for all $\din \in \Din$, 
$\bijD(I(\inp)(d,\din)) = I'(\inp)(\bijD(d),\bijI(\din))$;
\item
for all $d \in D$, $\bijO(I(\outp)(d)) = I'(\outp)(\bijD(d))$;
\item
for all $d \in D$ and $f \in \widetilde{F}$, 
$\bijD(I(f)(d)) = I'(\bijF(f))(\bijD(d))$;
\item
for all $d \in D$ and $p \in P$, 
$\bijB(I(p)(d)) = I'(\bijP(p))(\bijD(d))$.
\end{itemize}
\end{definition}

This definition differs from the definition of isomorphism of 
non-interactive proto-algorithms only by the addition of conditions 
concerning the interpretation of the special function symbols $\inp$ and 
$\outp$.

Interactive proto-algorithms may also be considered too concrete in a 
way not covered by isomorphism of interactive proto-algorithms.
This issue is addressed in Section~\ref{subsect-inter-algo-equiv} and 
leads there to the introduction of two other equivalence relations, just 
as in the case of non-interactive proto-algorithms.

\subsection{On Steps, Runs and What is Computed}
\label{subsect-inter-step-run-comp}

As with a non-interactive algorithm, the intuition is that an 
interactive proto-algorithm A is something that goes through states. 
In Section~\ref{subsect-inter-algo-proto}, it was informally explained 
how the state that it is in determines what the possible steps to a 
next state consist of and to what next states they lead.
The algorithmic step function $\astep_A$ that is defined below 
formalizes this.
The computational step function $\cstep_A$ that is also defined below is 
like the algorithmic step function $\astep_A$, but conceals the steps 
that consist of inspecting conditions.

\begin{definition}
\label{def-inter-algo-step-fnc}
Let $A = (\Sigma,G,\cI)$ be an interactive proto-algorithm, where 
$\Sigma = (F,P)$, $G = (V,E,\LBLv,\LBLe,l,r)$, and 
$\cI = (D,\Din,\Dout,I)$.
Then the \emph{algorithmic step function $\astep_A$ induced by $A$} is 
the smallest total function from $\cS_A$ to $\cP(\cS_A) \diff \emptyset$ 
such that for all $d,d' \in D$, $\din \in \Din$, $\dout \in \Dout$, and
$v,v' \in V$:
\begin{center}
\renewcommand{\arraystretch}{1}
\begin{tabular}{@{}r@{}l@{\,}l@{}} 
$\astep_A((\din,\botc,\bot))$ & ${} \ni (\bot,(v',d),\bot)$ \\
\multicolumn{3}{l}{\hspace*{8.3em}
if $l(r) = \ini$, $(r,v') \in E$, and $I(\ini)(\din) = d$,}
\\
$\astep_A((\bot,(v,d),\bot))$ & ${} \ni (\bot,(v',d'),\bot)$ \\ 
\multicolumn{3}{l}{\hspace*{8.3em}
if $l(v) \in \widetilde{F}$, $(v,v') \in E$, and $I(l(v))(d) = d'$,}
\\
$\astep_A((\bot,(v,d),\bot))$ & ${} \ni (\bot,(v',d),\bot)$ \\ 
\multicolumn{3}{l}{\hspace*{8.3em}
if $l(v) \in P$, $(v,v') \in E$, and $I(l(v))(d) = l((v,v'))$,}
\\
$\astep_A((\bot,(v,d),\bot))$ & ${} \ni (\din,(v',d),\dout)$ \\ 
\multicolumn{3}{l}{\hspace*{8.3em}
if $l(v) = \outp$, $(v,v') \in E$, and $I(l(v))(d) = \dout$,}
\\
$\astep_A((\din,(v,d),\dout))$ & ${} \ni (\bot,(v',d'),\bot)$ \\ 
\multicolumn{3}{l}{\hspace*{8.3em}
if $l(v) = \inp$, $(v,v') \in E$, and $I(l(v))(d,\din) = d'$,}
\\
$\astep_A((\bot,(v,d),\bot))$ & ${} \ni (\bot,\botc,\dout)$ \\ 
\multicolumn{3}{l}{\hspace*{8.3em}
if $l(v) = \fin$ and $I(\fin)(d) = \dout$,}
\\
$\astep_A((\bot,\botc,\dout))$ & ${} \ni (\bot,\botc,\dout)$
\end{tabular}
\end{center}
and the \emph{computational step function $\cstep_A$ induced by $A$} is 
the smallest total function from $\cS_A$ to $\cP(\cS_A) \diff \emptyset$ 
such that, for all $d,d' \in D$, $\din \in \Din$, $\dout \in \Dout$, 
$v,v' \in V$, and $s \in \cS_A$:
\begin{center}
\renewcommand{\arraystretch}{1}
\begin{tabular}{@{}r@{}l@{\,}l@{}} 
$\cstep_A((\din,\botc,\bot))$ & ${} \ni (\bot,(v',d),\bot)$ \\
\multicolumn{3}{l}{\hspace*{8.3em}
if $l(r) = \ini$, $(r,v') \in E$, and $I(\ini)(\din) = d$,}
\\
$\cstep_A((\bot,(v,d),\bot))$ & ${} \ni (\bot,(v',d'),\bot)$ \\ 
\multicolumn{3}{l}{\hspace*{8.3em}
if $l(v) \in \widetilde{F}$, $(v,v') \in E$, and $I(l(v))(d) = d'$,}
\\
$\cstep_A((\bot,(v,d),\bot))$ & ${} \ni s$ \\ 
\multicolumn{3}{l}{\hspace*{8.3em}
if $l(v) \in P$, $(v,v') \in E$, $I(l(v))(d) = l((v,v'))$, and} \\
\multicolumn{3}{l}{\hspace*{8.3em} \phantom{if}
$\cstep_A((\bot,(v',d),\bot)) \ni s$,}
\\
$\cstep_A((\bot,(v,d),\bot))$ & ${} \ni (\din,(v',d),\dout)$ \\ 
\multicolumn{3}{l}{\hspace*{8.3em}
if $l(v) = \outp$, $(v,v') \in E$, and $I(l(v))(d) = \dout$,}
\\
$\cstep_A((\din,(v,d),\dout))$ & ${} \ni (\bot,(v',d'),\bot)$ \\ 
\multicolumn{3}{l}{\hspace*{8.3em}
if $l(v) = \inp$, $(v,v') \in E$, and $I(l(v))(d,\din) = d'$,}
\\
$\cstep_A((\bot,(v,d),\bot))$ & ${} \ni (\bot,\botc,\dout)$ \\ 
\multicolumn{3}{l}{\hspace*{8.3em}
if $l(v) = \fin$ and $I(\fin)(d) = \dout$,}
\\
$\cstep_A((\bot,\botc,\dout))$ & ${} \ni (\bot,\botc,\dout)$.
\end{tabular}
\end{center}
\end{definition}

The definitions of the algorithmic and computational step functions  
induced by an interactive proto-algorithm differ from the definitions 
of the algorithmic and computational step functions induced by a 
non-interactive proto-algorithm only by the addition of conditions for 
internal states $(\bot,(v,d),\bot)$ with $l(v) = \outp$ and interaction 
states.

Below, we define what the possible runs of an interactive 
proto-algorithm on a sequence of input values are and what is computed 
by an interactive proto-algorithm.
To this end, we first give some auxiliary definitions.

\begin{definition}
\label{def-inter-algo-semi-runs-fnc}
Let $A = (\Sigma,G,\cI)$ be an interactive proto-algorithm.
Then the \emph{algorithmic semi-run set function $\asruns_A$ induced by 
$A$} and the \emph{computational semi-run set function $\csruns_A$ 
induced by $A$} are the total functions from $\cS_A$ to 
$\cP({\cS_A}^\infty)$ such that for all $s \in \cS_A$:
\begin{center}
\renewcommand{\arraystretch}{1}
\hspace*{1.5em}
\begin{tabular}{@{}l@{}l@{\;}l@{}} 
$\asruns_A(s)$ & ${} = 
 \set{\seq{s} \conc \stseq \where
      \Lexists{s' \in \astep_A(s)}{\stseq \in \asruns_A(s')}}$ &
if $s \notin \Sfin_A$,
\\
$\asruns_A(s)$ & ${} = \set{\seq{s}}$ & 
if $s \in \Sfin_A$.
\end{tabular}
\end{center} 
and
\begin{center}
\renewcommand{\arraystretch}{1}
\hspace*{1.5em}
\begin{tabular}{@{}l@{}l@{\;}l@{}} 
$\csruns_A(s)$ & ${} = 
 \set{\seq{s} \conc \stseq \where
      \Lexists{s' \in \cstep_A(s)}{\stseq \in \csruns_A(s')}}$ &
if $s \notin \Sfin_A$,
\\
$\csruns_A(s)$ & ${} = \set{\seq{s}}$ & 
if $s \in \Sfin_A$.
\end{tabular}
\end{center} 
A $\sigma \in {\cS_A}^\infty$ is called an \emph{algorithmic semi-run} 
if there exists an $s \in \cS_A$ such that $\sigma \in \asruns_A(s)$. 
\end{definition}
Henceforth, we write $\Asruns_A$, where $A$ is an interactive 
proto-algorithm, for the set of all algorithmic semi-runs.

The definitions of the algorithmic and computational semi-run set 
functions induced by an interactive proto-algorithm and the definitions 
of the algorithmic and computational semi-run set functions induced by 
a non-interactive proto-algorithm are practically the same.
The former definitions differ from the latter definitions only in that 
$\cS_A$ refers to the states of an interactive proto-algorithm rather 
than the states of a non-interactive proto-algorithm and that $\astep_A$ 
and $\cstep_A$ refer to step functions induced by an interactive 
proto-algorithm rather than step functions induced by a non-interactive 
proto-algorithm.

As with non-interactive proto-algorithms, when defining what is computed 
by an interactive proto-algorithm, a distinction must be made between 
convergent runs and divergent runs.
\begin{definition}
\label{def-inter-algo-div-conv}
Let $A = (\Sigma,G,\cI)$ be an interactive proto-algorithm, and 
let $\stseq \in \Asruns_A$. 
Then \emph{$\stseq$ is divergent} if there exists a suffix $\stseq'$ of 
$\stseq$ such that $\stseq' \in {\Sint_A}^\infty$, and 
\emph{$\stseq$ is convergent} if $\stseq$ is not divergent.
\end{definition}

The functions $\inputs_A$ and $\outputs_A$ defined below give the 
sequence of input values consumed by a given convergent algorithmic 
semi-run and the sequence of output values produced by a given 
convergent algorithmic semi-run, respectively.
\begin{definition}
\label{def-inter-algo-inputs-outputs-fnc}
Let $A = (\Sigma,G,\cI)$ be an interactive proto-algorithm.
Then the \emph{input value stream extraction function $\inputs_A$ 
for $A$} and the \emph{output value stream extraction function 
$\outputs_A$ for $A$} are the total functions from 
$\set{\sigma \in \Asruns_A \where \sigma \mathrm{\;is\;convergent}}$ to 
$\Din^\infty$ and $\Dout^\infty$, respectively, such that for all
$(\din,c,\dout) \in \cS_A$ and 
$\stseq \in \cS_A^\infty \union \set{\emptyseq}$ such that 
$\seq{(\din,c,\dout)} \conc \stseq \in \Asruns_A$ and
$\seq{(\din,c,\dout)} \conc \stseq$ is convergent: 
\begin{center}
\renewcommand{\arraystretch}{1}
\hspace*{1.5em}
\begin{tabular}{@{}l@{}l@{\;}l@{}} 
$\inputs_A(\seq{(\din,c,\dout)} \conc \stseq)$ & ${} = 
 \seq{\din} \conc \inputs_A(\stseq)$ &
if $\stseq \neq \emptyseq$ and $\din \neq \bot$,
\\
$\inputs_A(\seq{(\din,c,\dout)} \conc \stseq)$ & ${} = 
 \inputs_A(\stseq)$ &
if $\stseq \neq \emptyseq$ and $\din = \bot$,
\\
$\inputs_A(\seq{(\din,c,\dout)} \conc \stseq)$ & ${} = 
 \emptyseq$ &
if $\stseq = \emptyseq$
\end{tabular}
\end{center} 
and
\begin{center}
\renewcommand{\arraystretch}{1}
\hspace*{1.5em}
\begin{tabular}{@{}l@{}l@{\;}l@{}} 
$\outputs_A(\seq{(\din,c,\dout)} \conc \stseq)$ & ${} = 
 \seq{\dout} \conc \outputs_A(\stseq)$ &
if $\stseq \neq \emptyseq$ and $\dout \neq \bot$,
\\
$\outputs_A(\seq{(\din,c,\dout)} \conc \stseq)$ & ${} = 
 \outputs_A(\stseq)$ &
if $\stseq \neq \emptyseq$ and $\dout = \bot$,
\\
$\outputs_A(\seq{(\din,c,\dout)} \conc \stseq)$ & ${} = 
 \seq{\dout}$ &
if $\stseq = \emptyseq$.
\end{tabular}
\end{center} 
\end{definition}

As with non-interactive algorithms, a run of an interactive algorithm is 
a semi-run of the interactive algorithm concerned that starts in an 
initial state.
The following definition concerns what the possible runs of an 
interactive proto-algorithm on a sequence of input values are.
As with steps, a distinction is made between algorithmic runs and 
computational runs.
\begin{definition}
\label{def-inter-algo-runs}
Let $A = (\Sigma,G,\cI)$ be an interactive proto-algorithm, where
$\cI = (D,\Din,\Dout,I)$, and let $\dsin \in \Din^\infty$.
Then the \emph{algorithmic run set of $A$ on $\dsin$}, 
written $\arun_A(\dsin)$, 
is the set of all $\stseq \in \asruns_A((\dsin[1],\botc,\bot))$ such 
that $\inputs_A(\stseq) = \dsin$
and the \emph{computational run set of $A$ on $\dsin$}, 
written $\crun_A(\dsin)$, 
is the set of all $\stseq \in \csruns_A((\dsin[1],\botc,\bot))$ such 
that $\inputs_A(\stseq) = \dsin$.
\end{definition}

The following definition concerns what is computed by an interactive 
proto-algorithm.
\begin{definition}
\label{def-inter-algo-computed-rel}
Let $A = (\Sigma,G,\cI)$ be an interactive proto-algorithm, where 
$\cI = (D,\Din,\Dout,I)$.
Then the \emph{relation $\widehat{A}$ computed by $A$} is the relation 
from ${\Din}^\infty$ to ${\Dout}^\infty$ such that 
for all $\dsin \in \Din^\infty$ and $\dsout \in \Dout^\infty$:
\begin{center}
\renewcommand{\arraystretch}{1}
\begin{tabular}{@{}l@{}} 
$(\dsin,\dsout) \in \widehat{A}$ iff
there exists a convergent $\sigma \in \arun_A(\dsin)$ such that 
$\dsout = \outputs(\sigma)$.
\end{tabular}
\end{center}
\end{definition}

The following is a corollary of Definitions
\ref{def-noninter-algo-step-fnc}--\ref{def-noninter-algo-computed-rel} 
and 
\ref{def-inter-algo-step-fnc}--\ref{def-inter-algo-computed-rel}.
\begin{corollary}
\label{corollary-noninter-inter}
Let $A = (\Sigma,G,\cI)$ be a non-interactive proto-algorithm, where 
$\Sigma = (F,P)$, $G = (V,E,\LBLv,\LBLe,l,r)$, and 
$\cI = (D,\Din,\Dout,I)$.
Moreover, 
let $A' = (\Sigma',G',\cI')$ be an interactive proto-algorithm, 
where $\Sigma' = (F \union \set{\inp,\outp},P)$,
$G' = (V,E,\LBLv',\LBLe,l',r)$, where 
$\LBLv' = \LBLv \union \set{\inp,\outp}$ and
$l'$ is such that $l$ is $l'$ restricted to $\LBLv \union \LBLe$, and
$\cI' = (D,\Din,\Dout,I')$, where 
$I'$ is such that $I$ is $I'$ restricted to $\LBLv \union \LBLe$.
Then:
\begin{itemize}
\item 
for all $s,s' \in \cS_A$, 
$s' \in \astep_A(s)$ iff $s' \in \astep_{A'}(s)$;
\item
for all $\din \in \Din$ and $\dout \in \Dout$,
$(\din,\dout) \in \widehat{A}$ iff 
$(\seq{\din},\seq{\dout}) \in \widehat{A'}$.
\end{itemize}
\end{corollary}
Corollary~\ref{corollary-noninter-inter} justifies the statement that 
the notion of an interactive proto-algo\-rithm is a generalization of 
the notion of a non-interactive proto-algorithm.

\subsection{Algorithmic and Computational Equivalence}
\label{subsect-inter-algo-equiv}

As with non-interactive proto-algorithms, if an interactive 
proto-algorithm $A'$ can mimic an interactive proto-algorithm $A$ 
step-by-step, then we say that $A$ is algorithmically simulated by $A'$.
If the steps that consist of inspecting conditions are ignored, then we 
say that $A$ is computationally simulated by $A'$.
Algorithmic and computational simulation can be formally defined using 
the step functions defined in Section~\ref{subsect-inter-step-run-comp}.

\begin{definition}
\label{def-inter-algo-sim}
Let $A = (\Sigma,G,\cI)$ and $A' = (\Sigma',G',\cI')$ be two interactive
proto-algorithms, where
$\cI = (D,\Din,\Dout,I)$ and $\cI' = (D',\Din',\Dout',I')$.
Then an \emph{algorithmic simulation of $A$ by $A'$} is a set
$R \subseteq \cS_A \x \cS_{A'}$ such that: 
\begin{itemize}
\item
for all $s \in \cS_A$ and $s' \in \cS_{A'}$:
\begin{itemize}
\item
if $\sini \in \Sini_A$, then there exists an 
$\sini' \in \Sini_{A'}$ such that $(\sini,\sini') \in R$;
\item
if $\sfin' \in \Sfin_{A'}$, then there exists an  
$\sfin \in \Sfin_A$ such that $(\sfin,\sfin') \in R$;
\item
if $(s,s') \in R$ and $t \in \astep_A(s)$, 
then there exists a $t' \in \astep_{A'}(s')$ such that 
{$(t,t') \in R$};
\end{itemize}
\item
for all $(s,s') \in R$:
\begin{itemize}
\item
$s \in \Sini_A$ iff $s' = \Sini_{A'}$;
\item 
$s \in \Sfin_A$ iff $s' = \Sfin_{A'}$;
\item 
$s \in \Sint_A$ iff $s' = \Sint_{A'}$;
\end{itemize}
\item
there exist total functions $\funct{\fncI}{\Din}{\Din'}$ and 
$\funct{\fncO}{\Dout'}{\Dout}$ such that
for all $((\din,c,\dout),(\din',c',\dout')) \in R$:
\begin{itemize}
\item
if $\din \neq \bot$ or $\din' \neq \bot$, then $\din' = \fncI(\din)$;
\item 
if $\dout \neq \bot$ or $\dout' \neq \bot$, then $\dout = \fncO(\dout')$
\end{itemize}
\end{itemize}
and a \emph{computational simulation of $A$ by $A'$} is a set
$R \subseteq \cS_A \x \cS_{A'}$ such that: 
\begin{itemize}
\item
for all $s \in \cS_A$ and $s' \in \cS_{A'}$:
\begin{itemize}
\item
if $\sini \in \Sini_A$, then there exists an 
$\sini' \in \Sini_{A'}$ such that $(\sini,\sini') \in R$;
\item
if $\sfin' \in \Sfin_{A'}$, then there exists an  
$\sfin \in \Sfin_A$ such that $(\sfin,\sfin') \in R$;
\item
if $(s,s') \in R$ and $t \in \cstep_A(s)$, 
then there exists a $t' \in \cstep_{A'}(s')$ such that 
{$(t,t') \in R$};
\end{itemize}
\item
for all $(s,s') \in R$:
\begin{itemize}
\item
$s \in \Sini_A$ iff $s' = \Sini_{A'}$;
\item 
$s \in \Sfin_A$ iff $s' = \Sfin_{A'}$;
\item 
$s \in \Sint_A$ iff $s' = \Sint_{A'}$;
\end{itemize}
\item
there exist total functions $\funct{\fncI}{\Din}{\Din'}$ and 
$\funct{\fncO}{\Dout'}{\Dout}$ such that
for all $((\din,c,\dout),(\din',c',\dout')) \in R$:
\begin{itemize}
\item
if $\din \neq \bot$ or $\din' \neq \bot$, then $\din' = \fncI(\din)$;
\item 
if $\dout \neq \bot$ or $\dout' \neq \bot$, then $\dout = \fncO(\dout')$.
\end{itemize}
\end{itemize}
$A$ \emph{is algorithmically simulated by} $A'$, written $A \asim A'$,
if there exists an algorithmic simulation of $A$ by $A'$. 
\\ 
$A$ \emph{is computationally simulated by} $A'$, written $A \csim A'$, 
if there exists a computational simulation of $A$ by $A'$.
\\ 
$A$ \emph{is algorithmically equivalent to} $A'$, written $A \aeqv A'$, 
if there exist an algorithmic simulation $R$ of $A$ by $A'$ and an 
algorithmic simulation $R'$ of $A'$ by $A$ such that $R' = R^{-1}$.
\\
$A$ \emph{is computationally equivalent to} $A'$, written $A \ceqv A'$, 
if there exist a computational simulation $R$ of $A$ by $A'$ and a 
computational simulation $R'$ of $A'$ by $A$ such that $R' = R^{-1}$.
\end{definition}

The conditions imposed on an algorithmic or computational simulation $R$ 
of an interactive proto-algorithm $A$ by an interactive proto-algorithm 
$A'$ include, in addition to the usual transfer conditions, also 
conditions that guarantee that a state of $A$ is only related by $R$ to 
a state of $A'$ of the same kind (initial, final, internal or 
interaction) and conditions that guarantee that states of $A$ with the 
same input value are only related by $R$ to states of $A'$ with the same 
input value and that states of $A'$ with the same output value are only 
related by $R$ to states of $A$ with the same output value.
In the case of non-interactive proto-algorithms, the latter conditions 
are superfluous due to the lack of interaction.
Lemma~\ref{lemma-inter-simulation} given below, used in the proof of the 
theorem that follows it, would not hold without the additional 
conditions.

\begin{lemma}
\label{lemma-inter-simulation}
Let $A = (\Sigma,G,\cI)$ and $A' = (\Sigma',G',\cI')$ be interactive
proto-algorithms, where $\Sigma = (F,P)$, $\Sigma' = (F',P')$, 
$G = (V,E,\LBLv,\LBLe,l,r)$, $G' = (V',E',\LBLv',\LBLe',l',r')$, 
$\cI = (D,\Din,\Dout,I)$, and $\cI' = (D',\Din',\Dout',I')$, 
let $R \subseteq \cS_A \x \cS_{A'}$ and 
$\funct{\fncI}{\Din}{\Din'}$ be such that, for all $\din \in \Din$, 
$((\din,\botc,\bot),(\fncI(\din),\botc,\bot)) \in R$, and
let $\funct{\fncI^\infty}{\Din^\infty}{{\Din'}^\infty}$ be the pointwise 
extension of $\fncI$.
Then $R$ is an algorithmic simulation of $A$ by $A'$ only if,
for all $\dsin \in \Din^\infty$, for all $\sigma \in \arun_A(\dsin)$,
there exists a $\sigma' \in \arun_{A'}(\fncI^\infty(\dsin))$ such that,
for all $n \in \Natpos$, $(\sigma[n],\sigma'[n]) \in R$. 
\end{lemma}
\begin{proof}
Let $R$ be an algorithmic simulation of $A$ by $A'$ and 
$\funct{\fncI}{\Din}{\Din'}$ be such that, for all $\din \in \Din$, 
$((\din,\botc,\bot),(\fncI(\din),\botc,\bot)) \in R$, 
let $\dsin \in \Din^\infty$, and let $\sigma \in \arun_A(\dsin)$.
Then we can easily construct a 
$\sigma' \in \arun_{A'}(\fncI^\infty(\dsin))$ 
such that, for all $n \in \Natpos$, $(\sigma[n],\sigma'[n]) \in R$,
using the conditions imposed on algorithmic simulations. 
\qed
\end{proof}

The following theorem tells us that, 
if an interactive proto-algorithm $A$ is algorithmically simulated by an 
interactive proto-algorithm $A'$, then 
(a)~the relation computed by $A'$ models the relation computed by $A$ 
and
(b)~for each finite convergent algorithmic run of $A$, the simulation 
results in a finite convergent algorithmic run of $A'$ consisting of the 
same number of algorithmic steps.
\begin{theorem}
\label{theorem-inter-alg-equiv}
Let $A = (\Sigma,G,\cI)$ and $A' = (\Sigma',G',\cI')$ be interactive
proto-algorithms, where 
$\cI = (D,\Din,\Dout,I)$, and $\cI' = (D',\Din',\Dout',I')$.
Then $A \asim A'$ only if there exist total functions 
$\funct{\fncI}{\Din}{\Din'}$ and $\funct{\fncO}{\Dout'}{\Dout}$ 
such that: 
\begin{enumerate}
\item[(1)]
for all $\dsin \in \Din^\infty$ and $\dsout \in \Dout^\infty$,
$(\dsin,\dsout) \in \widehat{A}$ only if 
there exists a $\dsout' \in {\Dout'}^\infty$ such that
$(\fncI^\infty(\dsin),\dsout') \in \widehat{A'}$ and 
$\fncO^\infty(\dsout') = \dsout$;
\item[(2)]
for all finite $\dsin \in \Din^\infty$, 
for all convergent $\sigma \in \arun_A(\dsin)$, 
there exists a convergent $\sigma' \in \arun_{A'}(\fncI^\infty(\dsin))$ 
with $\fncO^\infty(\outputs_{A'}(\sigma')) = \outputs_A(\sigma)$ 
such that $|\sigma| = |\sigma'|$;
\end{enumerate}
where
$\fncI^\infty$ is the pointwise extension of $\fncI$ to a function 
from $\Din^\infty$ to ${\Din'}^\infty$ and
$\fncO^\infty$ is the pointwise extension of $\fncO$ to a function 
from ${\Dout'}^\infty$ to $\Dout^\infty$.
\end{theorem}
\begin{proof}
Because $A \asim A'$, there exists an algorithmic simulation of $A$ by 
$A'$.

Let $R$ be an algorithmic simulation of $A$ by $A'$,
let $\fncI$ be a function from $\Din$ to $\Din'$ such that, 
for all $\din \in \Din$, 
$((\din,\botc,\bot),(\fncI(\din),\botc,\bot)) \in R$, and
let $\fncO$ be a function from $\Dout'$ to $\Dout$ such that, 
for all $\dout' \in \Dout'$, 
$((\bot,\botc,\fncO(\dout')),(\bot,\botc,\dout')) \in R$.
Functions $\fncI$ and $\fncO$ exist by the definition of an algorithmic
simulation.
By Lemma~\ref{lemma-inter-simulation}, 
for all $\dsin \in \Din^\infty$, for all $\sigma \in \arun_A(\dsin)$,
there exists a $\sigma' \in \arun_{A'}(\fncI^\infty(\dsin))$ such that,
for all $n \in \Natpos$, $(\sigma[n],\sigma'[n]) \in R$.

Let $\dsin \in \Din^\infty$, and
let $\sigma \in \arun_A(\dsin)$ and 
$\sigma' \in \arun_{A'}(\fncI^\infty(\dsin))$ 
be such that, for all $n \in \Natpos$, $(\sigma[n],\sigma'[n]) \in R$.
Moreover, assume that $G = (V,E,\LBLv,\LBLe,l,r)$ and 
$G' = (V',E',\LBLv',\LBLe',l',r')$.
Then from the definition of an algorithmic simulation, it immediately 
follows that, for all $n \in \Natpos$:
\begin{enumerate} 
\item[(a)]
$\sigma[n] \notin \Sint_A$ only if $\sigma[n] \notin \Sint_{A'}$; 
\item[(b)]
for all $\din \in \Din$, $\dout \in \Dout$, and $c \in V \x D$:
\begin{itemize}
\item
$\sigma[n] = (\din,\botc,\bot)$ only if 
$\sigma'[n] = (\fncI(\din),\botc,\bot)$;
\item
$\sigma[n] = (\bot,\botc,\dout)$ only if 
there exists a $\dout' \in \Dout'$ such that 
$\sigma'[n] = (\bot,\botc,\dout')$ and $\fncO(\dout') = \dout$;
\item   
$\sigma[n] = (\din,c,\dout)$ only if 
there exist a $c' \in V' \x D'$ and a $\dout' \in \Dout'$ such that 
$\sigma'[n] = (\fncI(\din),c',\dout')$, and $\fncO(\dout') = \dout$.
\end{itemize}
\end{enumerate}
By the definition of the relation computed by an interactive 
proto-algorithm, both~(1) and~(2) follow immediately from~(a) and~(b).
\qed
\end{proof}
It is easy to see that Theorem~\ref{theorem-inter-alg-equiv} goes 
through as far as (1) is concerned if algorithmic simulation is replaced 
by computational simulation.
However, (2) does not go through if algorithmic simulation is replaced 
by computational simulation.

Lemma~\ref{lemma-inter-simulation} and 
Theorem~\ref{theorem-inter-alg-equiv} are the counterparts of 
Lemma~\ref{lemma-noninter-simulation} and 
Theorem~\ref{theorem-noninter-alg-equiv} for interactive 
proto-algorithms.
The proofs of the former two results go along the same lines as the 
proofs of the latter two results.

The following theorem tells us how isomorphism, algorithmic equivalence, 
and computational equivalence are related.
\begin{theorem}
\label{theorem-inter-equivs}
Let $A$ and $A'$ be interactive proto-algorithms.
Then: 
\begin{trivlist}
\item[]
$\qquad$ (1) $\;$ $A \iso A'$ only if $A \aeqv A'$ 
$\qquad$ (2) $\;$ $A \aeqv A'$ only if $A \ceqv A'$.
\end{trivlist}
\end{theorem}
\begin{proof}
This is proved in the same way as Theorem~\ref{theorem-noninter-equivs}.
\qed
\end{proof}

As with non-interactive proto-algorithms, the opposite implications do 
not hold in general.
That is, there exist interactive proto-algorithms $A$ and $A'$ for which 
it does not hold that $A \iso A'$ if $A \aeqv A'$ and there exist 
interactive proto-algorithms $A$ and $A'$ for which it does not hold 
that $A \aeqv A'$ if $A \ceqv A'$.
In both cases, the construction of a general illustrating example for 
non-interactive proto-algorithms described in~\cite{Mid24a} also works 
for interactive proto-algorithms.

The remarks made about algorithmic equivalence and computational 
equivalence of non-interactive proto-algorithms in the last paragraph of 
Section~\ref{subsect-noninter-algo-equiv} also apply to algorithmic 
equivalence and computational equivalence of interactive 
proto-algorithms.

\section{Non-Interaction versus Interaction in Algorithms}
\label{sect-versus}

In order to discuss the similarities and differences between
non-interactive algorithms and interactive algorithms, it is useful to
distinguish between trivial interactive algorithms and non-trivial 
interactive algorithms and to take expansions of non-interactive 
algorithms into consideration.
\begin{definition}
\label{def-triv-inter-algo-proto}
Let $A = (\Sigma,G,\cI)$ be an interactive proto-algorithm, where 
$G = (V,E,\LBLv,\LBLe,l,r)$.
Then $A$ is a \emph{trivial interactive proto-algorithm} if, for all
$v \in V$, $l(v) \notin \set{\inp,\outp}$, and $A$ is a 
\emph{non-trivial interactive proto-algorithm} if $A$ is not a
trivial interactive proto-algorithm.
\end{definition}
\begin{definition}
\label{def-expand-algo-proto}
Let $A = (\Sigma,G,\cI)$ and $A' = (\Sigma',G',\cI')$ be two 
non-interactive proto-algorithms or two interactive proto-algorithm,
where $\Sigma = (F,P)$, $\Sigma' = (F',P')$, $\cI = (D,\Din,\Dout,I)$,
and $\cI' = (D',\Din',\Dout',I')$.
Then $A'$ is an \emph{expansion of} $A$ if $F \subseteq F'$, 
$P \subseteq P'$, $G = G'$, and $I$ is the restriction of $I'$ to
$F \union P$.
\end{definition}

A corollary of Definition~\ref{def-triv-inter-algo-proto} is that each 
trivial interactive algorithm is also a non-interactive algorithm. 
\begin{corollary}
\label{corollary-triv-inter-algo-proto}
Let $A$ be an interactive proto-algorithm.
Then $A$ is a trivial interactive proto-algorithm only if $A$ is a
non-interactive proto-algorithm.
\end{corollary}

The alphabet of each interactive algorithm includes the symbols $\inp$ 
and $\outp$.
Therefore, we do not have that each non-interactive algorithm is also 
a trivial interactive algorithm.
However, a corollary of Definitions~\ref{def-triv-inter-algo-proto} 
and~\ref{def-expand-algo-proto} is that each non-interactive algorithm 
can be expanded to a trivial interactive algorithm.
\begin{corollary}
\label{corollary-expand-algo-proto}
Let $A$ be a non-interactive proto-algorithm.
Then there exists a trivial interactive proto-algorithm $A'$ such that
$A'$ is an expansion of $A$.
\end{corollary}

It immediately follows from the definitions of the notions of 
non-interactive and interactive proto-algorithms that 
Corollaries~\ref{corollary-triv-inter-algo-proto} 
and~\ref{corollary-expand-algo-proto} do not go through if the 
restriction to trivial interactive proto-algorithms is dropped.
The first question that remains is whether each non-trivial interactive 
proto-algorithm is algorithmically equivalent to a trivial interactive 
proto-algorithm.
If this question is to be answered in the negative, then the next
question is whether each non-trivial interactive proto-algorithm is
algorithmically simulated by a trivial interactive proto-algorithm.
Below I argue that both questions must be answered in the negative, 
even if the condition that an input domain must be a finitely generated 
set is dropped.

In order to simulate a non-trivial interactive proto-algorithm by a 
trivial interactive proto-algorithm, each possibly infinite sequence of 
input values of the former must be treated as a single input value of 
the latter. 
The latter must be such that the interpretation of $\ini$ applied to a 
sequence of input values of the former yields an element of its 
algorithm domain that includes a representation of the sequence of input 
values concerned and a next input value can be obtained by applying a 
unary function on the algorithm domain.

It is easy to see that this leads to a trivial interactive 
proto-algorithm with an input domain that is not a finitely generated 
set and to a choice structure that differs from the choice structure of 
the non-trivial interactive proto-algorithm.
Irrespective of whether the condition that an input domain must be a 
finitely generated set is dropped, the different choice structures rule 
out algorithmic equivalence of the non-trivial interactive 
proto-algorithm and the trivial interactive proto-algorithm.

By its definition, a trivial interactive proto-algorithm is an 
interactive proto-algorithm without interaction states.
It follows immediately from the definition of algorithmic simulation 
that the absence of interaction states rules out algorithmic simulation
of a non-trivial interactive proto-algorithm by a trivial interactive 
proto-algorithm, again irrespective of whether the condition that an 
input domain must be a finitely generated set is dropped.
I conjecture that each non-trivial interactive proto-algorithm is
algorithmically simulated by a trivial interactive proto-algorithm if 
the definition of an algorithmic simulation for interactive 
proto-algorithms is weakened such that interaction states may also be 
related to internal states.

\newpage

\section{Concluding Remarks}
\label{sect-conclusions}

In~\cite{Mid24a} an account is given of a quest for a satisfactory 
formalization of the notion of an algorithm that is informally 
characterized in standard works from the mathematical and computer 
science literature.
That notion only covers algorithms that are deterministic and 
non-interactive.
In this paper, the quest is extended to non-deterministic and 
interactive algorithms.

Non-deterministic algorithms, introduced in~\cite{Flo67a}, are widely 
used as a starting point for developing deterministic backtracking 
algorithms and play an important role in the field of computational 
complexity.
In this paper, an attempt has been made to generalize the results 
of the above-mentioned quest to a notion of an algorithm that covers 
both deterministic and non-deterministic algorithms that are 
non-interactive.
The term non-interactive algorithm has been coined for this notion of an
algorithm.

Interactive algorithms, which emerged due to the advent of interactive 
computation, are currently widespread.
In this paper, an attempt has been made to generalize the results 
of the above-mentioned quest further to a notion of an algorithm that 
covers both deterministic and non-deterministic algorithms that are 
interactive.
The main goals of this work are to provide a framework for studying 
complexity-theoretic issues concerning interactive algorithms and to
provide a semantic basis for languages to describe interactive 
algorithms.

A contemporary notion of an algorithm that is not treated in this paper,
is the notion of a concurrent algorithm.
Concurrent algorithms emerged due to the advent of parallel computation.
They can be briefly described as follows: ``The pattern of behaviour 
expressed by a concurrent algorithm consists of multiple parts that can 
take place concurrently''.
Until now I failed to generalize the results from~\cite{Mid24a} to a 
satisfactory general notion of a concurrent algorithm.
My thesis is that a satisfactory notion of a concurrent algorithm will 
be such that each concurrent proto-algorithm is algorithmically 
equivalent to a concurrent proto-algorithm that consist of only one 
part.


\bibliographystyle{splncs04}
\bibliography{ALG}

\end{document}